\documentclass[journal]{IEEEtran}
\usepackage{amsmath,amsfonts,amssymb}
\usepackage{algpseudocode}
\usepackage{algorithm}
\usepackage{array}
\usepackage{subfigure}
\usepackage{textcomp}
\usepackage{stfloats}
\usepackage{url}
\usepackage{verbatim}
\usepackage{graphicx}
\def\BibTeX{{\rm B\kern-.05em{\sc i\kern-.025em b}\kern-.08em
    T\kern-.1667em\lower.7ex\hbox{E}\kern-.125emX}}
\usepackage{balance}

\usepackage{cite}
\usepackage{makecell}
\usepackage{adjustbox}

\begin{document}
\algblock{ParFor}{EndParFor}

\algnewcommand\algorithmicparfor{\textbf{parfor\:}}
\algnewcommand\algorithmicpardo{\textbf{do}}
\algnewcommand\algorithmicendparfor{\textbf{end parfor}}

\algrenewtext{ParFor}[1]{\algorithmicparfor #1 \algorithmicpardo}
\algrenewtext{EndParFor}{\algorithmicendparfor}

\title{Priority-based Energy Allocation in Buildings through Distributed Model Predictive Control}

\author{Hongyi Li, Jun Xu,~\IEEEmembership{Senior Member,~IEEE}, and Qianchuan Zhao,~\IEEEmembership{Senior Member,~IEEE}
\thanks{Corresponding author: Jun Xu (e-mail: xujunqgy@hit.edu.cn)}}

\markboth{IEEE TRANSACTIONS ON AUTOMATION SCIENCE AND ENGINEERING}%
{How to Use the IEEEtran \LaTeX \ Templates}

\maketitle

\begin{abstract}
Many countries are facing energy shortages today and most of the global energy is consumed by HVAC systems in buildings. For the scenarios where the energy system is not sufficiently supplied to HVAC systems, a priority-based allocation scheme based on distributed model predictive control is proposed in this paper, which distributes the energy rationally based on priority order. According to the scenarios, two distributed allocation strategies, i.e., one-to-one priority strategy and multi-to-one priority strategy, are developed in this paper and validated by simulation in a building containing three zones and a building containing 36 rooms, respectively. Both priority-based strategies fully exploit the potential of predictive control solutions.
The experiment shows that our scheme has good scalability and achieves the performance of the centralized strategy while making the calculation tractable.
\end{abstract}

\def\abstractname{Note to Practitioners}
\begin{abstract}
 The motivation of this paper is to develop a priority-based allocation strategy adapted to energy-limited systems. When energy is limited, the strategy can rationally allocate energy and satisfy the urgent need for energy supply in some specific zones. Two priority strategies are proposed for the case that a single subsystem corresponds to a particular priority and multiple subsystems correspond to the same priority, respectively.
 The developed strategies have been validated by co-simulation with MATLAB and EnergyPlus in a small-scale three-zone building and a large-scale 36-zone building to show their effectiveness.
\end{abstract}


\section{Introduction}
The building sector accounts for more than 40\% of global energy consumption \cite{espejel2022hvac,park2019development}, and more than 1/3 of carbon emissions are contributed by this sector \cite{sun2022carbon,too2022framework}. Under the United Nations Framework Convention on Climate Change (UNFCCC), China has committed to peak CO$_2$ emissions with a target date of 2030, announcing that it will reduce CO$_2$ emissions per unit of gross domestic product by 60–65\% of the emission levels in 2005 \cite{yang2023global,ding2019exploring}. 
Heating, Ventilation and Air Conditioning (HVAC) systems are the main consumers of energy in buildings, which makes them suitable candidates for improving building energy efficiency and reducing energy consumption, thus contributing to limiting global carbon emissions. Well-designed control rules applied to the building's  HVAC module offer a promising approach to improve the building's  energy efficiency.

Due to the advantages of low cost, simple operation and easy implementation, traditional control methods are still used in a large number of buildings to control HVAC systems \cite{afram2014theory,rawlings2018economic}, including manual control, feedback control, feed-forward control, PID control and so on. However, the control parameters of traditional methods are difficult to adjust and cannot handle the nonlinear dynamics of the system, which often leads to overshooting and makes the building operation inefficient. \cite{gupta2023optimal} proposes a hybrid particle swarm optimization algorithm for optimal tuning of conventional PID controllers and compares this method with the Ziegler-Nichols method, but this method is still not able to deal with the nonlinear dynamics and disturbances of the HVAC system.
In order to deal with the nonlinear dynamics of the building, rule-based control methods have been developed, which are based on a series of ``if-then-else" rules written based on expert experience or a priori knowledge to control the HVAC system. \cite{saloux2020optimal,alimohammadisagvand2018comparison} have achieved good control results using rule-based control methods. Although these rule-based methods can take into account the real-time constraints and nonlinearities of the system and do improve the energy efficiency of the system, a global optimization of the system is lacking and the rules are complex to write and maintain. As a result, these approaches require a great deal of experience and expertise in practical implementation and are difficult to generalize and replicate.

In recent years, model predictive control (MPC) has received considerable attention from researchers as a promising advanced control method. In addition to the well-known advantages of dealing with nonlinear dynamic models and constraints in a systematic mode, MPC controllers could take into account weather forecasts, room occupancy, and other information that may be relevant to the optimal control law of the system during operation. However, as the scale of the building complex increases, the complexity of the system's dynamic model increases. In this case, the number of decision variables to be processed by the MPC controller during each execution becomes very large, and limited computational power makes it difficult to deploy centralized MPC controllers in large-scale building systems.

Large-scale, multi-coupled and multi-constraint complex systems can be effectively handled by distributed model predictive control (DMPC), which is a suitable method for managing energy distribution in buildings, especially when the number of control variables and signals from sensors grows rapidly with the scale of the HVAC system. In addition to building sector, its application areas involve other engineering systems such as traffic control \cite{eini2019distributed}, smart grids \cite{del2014combined}, water supply systems \cite{zhang2019enhancing}, supply chains \cite{fu2019cooperative}, floating object transport \cite{chen2019distributed}, unmanned aerial vehicle formations \cite{cai2018formation}, vehicle formations \cite{bono2021swarm}, etc.

Distributed MPC splits the original centralized problem into multiple local subproblems, which are small and easy to solve. Coordination of the local subproblems can be performed locally through communication between the subproblems or through a global coordinator.
The distributed control technique is not significantly affected by the number of HVAC systems, so many scholars have started to investigate distributed techniques applied to large-scale buildings.
\cite{lefebure2022distributed} proposes a method based on dual decomposition for co-optimizing buildings and energy hubs to improve energy efficiency and conservation.
A distributed MPC framework based on Benders decomposition is illustrated in \cite{morocsan2011distributed} for controlling a building heating system consisting of a central heat source and multiple local heat sources. 
A limited communication DMPC algorithm for coupled and constrained linear discrete systems is proposed in \cite{jalal2016limited}. 
\cite{bay2022distributed} invokes the new concept of grid aggregator in the limited communication DMPC structure and extends the algorithm to enable buildings to interact with the grid.
Nash equilibrium and alternating direction method of multipliers (ADMM) is used in \cite{mork2022nonlinear} to coordinate the control inputs in each zone. 

In this paper, we propose a priority-based distributed scheme to rationally allocate building energy, which skillfully decouples the subsystems. The advantage of this approach is that instead of solving the global optimization problem, the optimization problem containing only the subsystem's own objective function and constraints is solved in parallel, and the proposed method can approach the performance of the centralized approach. Furthermore, the proposed distributed approach has the potential to be applied to more distributed settings besides the HVAC systems involved in this paper, such as demand management systems in microgrids \cite{korkas2018grid} and load management programs with smart zoning \cite{baldi2018automating}.

The contributions of this paper are as follows.
\begin{itemize}
	\item 
	For a limited amount of energy, it is not possible to make all the rooms satisfy the comfort requirements. The contribution of this paper is to provide a distributed scheme that allocates energy according to priority to satisfy the energy supply of the rooms that are in urgent need of energy supply. The algorithm is independent of the size of the system and is scalable. Optimization operations could be performed by subsystems in parallel.
	
	\item Priority-based allocation algorithm proposed in this paper exploits the potential of predictive control solutions, i.e., it utilizes all the solutions obtained by predictive control. In contrast, the original predictive control would only use the first element of the solution sequence and discard the other elements.

	\item Few literature has considered large-scale cases, the algorithm proposed in this paper is applied to consider a large-scale energy distribution scenario for 36 zones. Simulation experiments illustrate that the proposed algorithm is well suited for application in large-scale scenarios.

\end{itemize}

\section{System description}
\subsection{Overview}
\begin{figure}[htbp]
	\centering
	\includegraphics[width=0.32\textwidth]{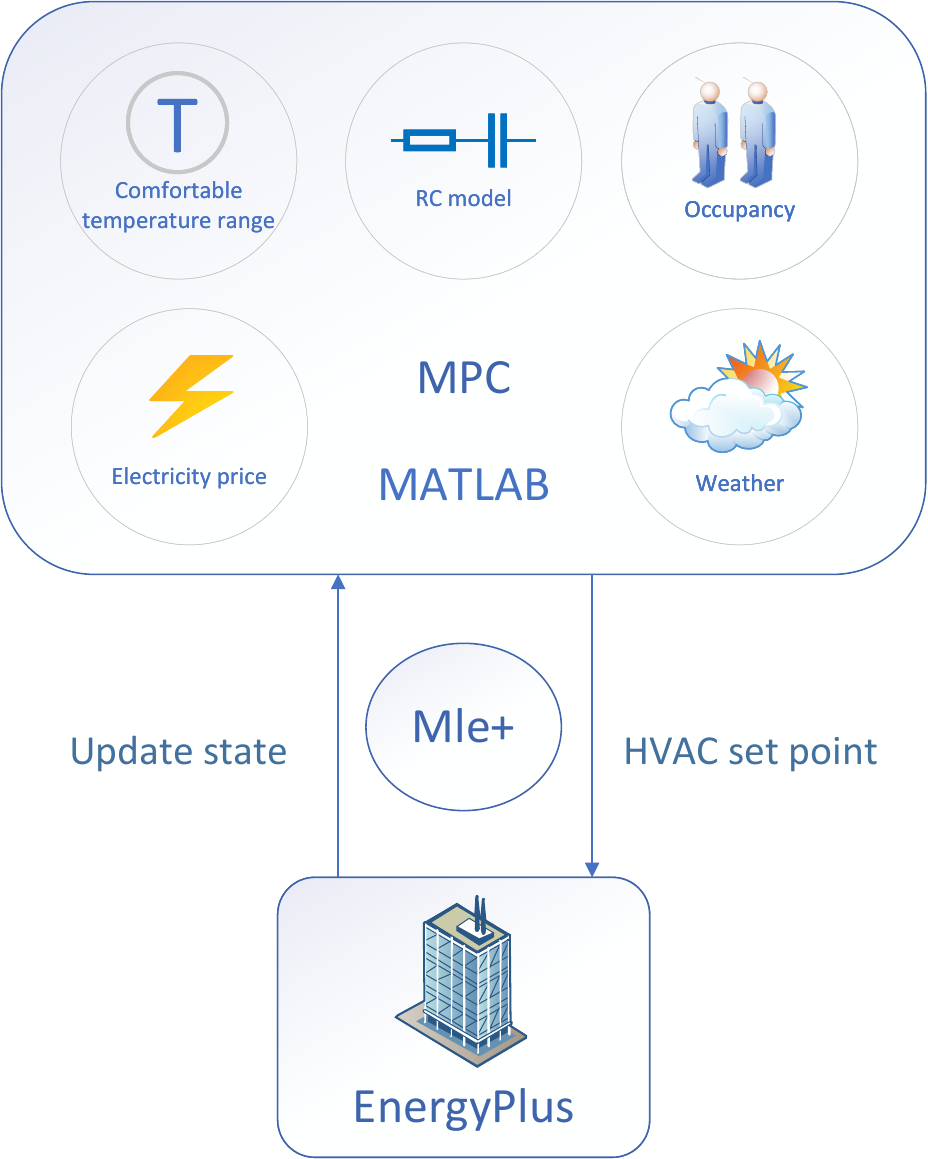} 
	\caption{Framework of energy allocation.} 
	\label{general_framework_priority} 
\end{figure}

The work in this paper is to rationally allocate energy to the zones based on MPC according to the priority order. Fig. \ref{general_framework_priority} shows the overall framework of the scheme in this paper. The control algorithm is deployed in MATLAB and the resistance-capacitance (RC) model is employed as the prediction model for predictive control. An optimization operation that considers building occupancy information, weather forecasts and local electricity prices yields the HVAC set point. The control action is then applied via Mle+ \cite{zhao2013energyplus} to the building model developed in EnergyPlus \cite{crawley2001energyplus}. After each optimization operation, the real-time states of the building in EnergyPlus are updated to MATLAB via Mle+ for the next optimization implementation.

\subsection{Prediction model for MPC}
The RC model is a commonly used prediction model in the building sector, which is built based on indoor nodes and wall nodes. Notations of the quantities in this section are shown in Table \ref{tab2}.
\begin{table}[htbp]
	\centering
	\caption{Notations of the quantities}
	\label{tab2}
	\begin{tabular}{cc}
		\hline
		Notation&Description \\
		\hline
        $u$& Input of the HVAC system\\
        $y$& Output of the HVAC system\\
        $x$& State of the HVAC system\\
		$ori$&\makecell[c]{Take the values $n$, $e$, $w$ and $s$,\\representing north, east, west, and south}\\
		$C_{z}$&Thermal capacity of the zone\\
		$C_{ {ori }}^{w}$&Thermal capacity of the wall\\
		$R_{ori}^w$&Thermal resistance for conduction of the wall\\
		$R_{ori}$&Thermal resistance for convection on inside surface\\
		$R_{ori}^\prime$&Thermal resistance for convection on outside surface\\
		$T_{z}$&Indoor air temperature\\
		$T_{ori}^{out}$&Outside temperature adjacent to the $ori$ wall\\
		$T_{ori2}^{out}$&Outside temperature adjacent to the $ori2$ wall\\
		$T_{ori}^{wi}$&Inside surface temperature of the $ori$ wall \\
		$T_{ori}^{wo}$&Outside surface temperature of the $ori$ wall \\
		$T_{out}$&Outdoor ambient temperature \\
		$\dot{Q}_{ori}^{rad}$ &  Solar radiation, $ori$ wall\\
		$\dot{Q}_{ori2}^{rad}$ &  Solar radiation, $ori2$ wall\\
		$\dot{Q}_{z}^{in}$ & Internal gains in the zone\\
		$\dot{Q}_{z}^{rad}$ & Solar radiation in the zone\\
		\hline
	\end{tabular}
\end{table}

For the zone, the differential equation about the indoor node is
\begin{equation}
	\begin{aligned}
		C_{z}^{} \frac{d T_{z}^{}}{d t}=&\frac{T_{n}^{w i}-T_{z}^{}}{R_{n}}+\frac{T_{e}^{w i}-T_{z}^{}}{R_{e}}+\frac{T_{w}^{w i}-T_{z}^{}}{R_{w}}\\
		&+\frac{T_{s}^{w i}-T_{z}^{}}{R_{s}}
		+u_{}+\dot{Q}_{z}^{i n}+\dot{Q}_{z}^{rad}
	\end{aligned}.
         \label{rc1}
\end{equation}
The differential equation for the inside surface of the wall in the $ori$ direction is
\begin{equation}
	C_{ {ori }}^{w} \frac{d T_{ {ori }}^{w i}}{d t}=\frac{T_{z}^{}-T_{ {ori }}^{w i}}{R_{ {ori }}}+\frac{T_{{ori }}^{w o}-T_{ {ori }}^{w i}}{R_{ {ori }}^{w}}.
 \label{rc2}
\end{equation}
The differential equation for the outside surface of the wall in the $ori$ direction is
\begin{equation}
	C_{ {ori }}^{w} \frac{d T_{ {ori }}^{w o}}{d t}=\frac{T_{ {ori }}^{out}-T_{ {ori }}^{w o}}{R_{ {ori }}^{\prime}}+\frac{T_{ {ori }}^{w i}-T_{ {ori }}^{w o}}{R_{ {ori }}^{w}}+\dot{Q}_{ {ori }}^{ {rad }}.
 \label{rc3}
\end{equation}

\begin{figure*}
\begin{center}
    \adjustbox{valign=c,rotate=0}{%
    \begin{minipage}{\textwidth}
        \begin{equation}
        \bar{A}=\left[\begin{array}{ccccccccc}
        -\frac{1}{C_{z}} S & \frac{1}{C_{z} R_{n}} & \frac{1}{C_{z} R_{e}} & \frac{1}{C_{z} R_{w}} & \frac{1}{C_{z} R_{s}}  &0&0&0&0\\
        \frac{1}{C_{n}^{w} R_{n}} & \frac{-R_{n}-R_{n}^{w}}{C_{n}^{w} R_{n} R_{n}^{w}} & 0 & 0 & 0 &\frac{1}{C_{n}^{w} R_{n}^{w}}&0&0&0\\
        \frac{1}{C_{e}^{w} R_{e}} & 0 &  \frac{-R_{e}-R_{e}^{w}}{C_{e}^{w} R_{e} R_{e}^{w}} & 0 & 0 &0&\frac{1}{C_{e}^{w} R_{e}^{w}}&0&0\\
        \frac{1}{C_{w}^{w} R_{w}} & 0 & 0 &  \frac{-R_{w}-R_{w}^{w}}{C_{w}^{w} R_{w} R_{w}^{w}} & 0 &0&0&\frac{1}{C_{w}^{w} R_{w}^{w}}&0\\
        \frac{1}{C_{s}^{w} R_{s}} & 0 & 0 & 0 &  \frac{-R_{s}-R_{s}^{w}}{C_{s}^{w} R_{s} R_{s}^{w}} &0&0&0&\frac{1}{C_{s}^{w} R_{s}^{w}}\\
        0 & \frac{1}{C_{n}^{w} R_{n}^{w}} & 0 & 0 & 0 & \frac{-R_{n}^{\prime}-R_{n}^{w}}{C_{n}^{w} R_{n}^{\prime} R_{n}^{w}}&0&0&0\\
        0 & 0 & \frac{1}{C_{e}^{w} R_{e}^{w}} & 0 & 0 &0& \frac{-R_{e}^{\prime}-R_{e}^{w}}{C_{e}^{w} R_{e}^{\prime} R_{e}^{w}}&0&0\\
        0 & 0 & 0 & \frac{1}{C_{w}^{w} R_{w}^{w}} & 0 &0&0& \frac{-R_{w}^{\prime}-R_{w}^{w}}{C_{w}^{w} R_{w}^{\prime} R_{w}^{w}}&0\\
        0 & 0 & 0 & 0 & \frac{1}{C_{s}^{w} R_{s}^{w}} &0&0&0& \frac{-R_{s}^{\prime}-R_{s}^{w}}{C_{s}^{w} R_{s}^{\prime} R_{s}^{w}}\\
        \end{array}\right]
        \label{AA}
        \end{equation}
    \end{minipage}}
    \adjustbox{valign=c,rotate=0}{%
    \begin{minipage}{\textwidth}
        \begin{equation}
        \begin{array}{lllllllll}
	\bar{B}=[\frac{1}{C_{z}}&0 & 0 & 0&0&0&0&0&0]^\intercal
        \end{array}
        \label{BB}
        \end{equation}
    \end{minipage}}
    \adjustbox{valign=c,rotate=0}{%
    \begin{minipage}{\textwidth}
        \begin{equation}
        \begin{array}{lllllllll}
	\bar{d}=[\frac{\dot{Q}_{z}^{i n}+\dot{Q}_{z}^{rad}}{C_{z}}&0&0&0&0
 &\frac{T_n^{out}}{C_{n}^{w} R_{n}^{\prime}}+\frac{\dot{Q}_{n}^{rad}}{C_{n}^{w}}
 &\frac{T_e^{out}}{C_{e}^{w} R_{e}^{\prime}}+\frac{\dot{Q}_{e}^{rad}}{C_{e}^{w}}
 &\frac{T_w^{out}}{C_{w}^{w} R_{w}^{\prime}}+\frac{\dot{Q}_{w}^{rad}}{C_{w}^{w}}
 &\frac{T_s^{out}}{C_{s}^{w} R_{s}^{\prime}}+\frac{\dot{Q}_{s}^{rad}}{C_{s}^{w}}] ^\intercal
\end{array}
\label{dd}
        \end{equation}
    \end{minipage}}
\end{center}
\end{figure*}

For one zone with four walls, the indoor node can be modeled by the differential equation shown in (\ref{rc1}), and the inside surface and outside surface of four walls can be modeled by (\ref{rc2}) and (\ref{rc3}), respectively. Further, the state variable of the system is composed of the indoor air temperature in the zone and the inside and outside surface temperatures of the four oriented walls in the zone, i.e.,
$
x=[\begin{array}{ll}
	T_{z} & T_{n}^{wi} 
\end{array}
$
$
\begin{array}{lllllll}
	T_{e}^{wi} & T_{w}^{wi} & T_{s}^{wi}& T_{n}^{wo} & T_{e}^{wo} & T_{w}^{wo} & T_{s}^{wo}
\end{array}]^\intercal
$
$
\in \mathbb{R}^{9}
$.
The system input is $u\in \mathrm{R}$. We can obtain the RC model of one zone, containing nine differential equations, and the RC expression for one zone is
\begin{equation}
		{x}^+ =\bar{A} x + \bar{B} u +  \bar{d},\\
	\label{state2}
\end{equation}
where $\bar{A}$,$\bar{B}$ and $\bar{d}$ are given in (\ref{AA}), (\ref{BB}) and (\ref{dd}), respectively, and $S=\frac{1}{R_{n}}+\frac{1}{R_{e}}+\frac{1}{R_{w}}+\frac{1}{R_{s}}$ in (\ref{AA}).

Therefore, the RC model for the $m$-th zone in a multi-zone system can be modeled and discretized as
\begin{equation}
	\begin{aligned}
		{x}_m(k+1) &=A_m x_m(k) + B_m u_m(k) +  d_m(k)\\
		y_m(k) &= C_mx(k)
	\end{aligned}.
	\label{state_m}
\end{equation}
$A_m$, $B_m$, and $d_m$ are the corresponding matrices after discretisation.
$y_m$ is the output of the $m$-th zone and represents the indoor air temperature, and
$
\begin{array}{lllllllll}
	C_m=[ 1&0 & 0 & 0&0&0&0&0&0]
\end{array}\in \mathbb{R}^{9}
$. 

We can obtain the RC model for multiple zones in the same way. The RC model of multi-zone is established and discretized as
\begin{equation}
	\begin{aligned}
		{x}(k+1) &=A x(k) + B u(k) +  d(k)\\
		y(k) &= Cx(k)
	\end{aligned},
	\label{state}
\end{equation}
where $x=[x^\intercal_1,\:\cdots\:x^\intercal_N]^\intercal$, $u=[u^\intercal_1,\:\cdots\:u^\intercal_N]^\intercal$, $d=[d^\intercal_1,\:\cdots\:d^\intercal_N]^\intercal$, and $N$ is the number of zones.
$A$, $B$, and $d$ are the corresponding matrices about the whole system after discretisation.

The R-values and C-values in this paper are shown in Table \ref{rc}. For validation of the RC model and more details see \cite{li2023economic} and \cite{haghighi2013controlling}.
\begin{table}[htbp]
	\centering
	\caption{R-values and C-values}
	\label{rc}
	\begin{tabular}{cccc}
		\hline
		Parameter&Value($J/K$)&Parameter&Value($K/W$) \\
		\hline
		$R_{e,w}$ & 0.0232 & $C_{z}$ & $4.8\times 10^4$\\
		$R_{n,s}$ & 0.0310 & $C_{ {n}}^{w}$ & $8.5\times 10^5$\\
		$R_{e,w}^w$ & 0.0179 & $C_{ {e}}^{w}$ & $1.1\times 10^6$\\
		$R_{n,s}^w$ & 0.0238 & $C_{ {w}}^{w}$ & $1.1\times 10^6$\\
		$R_{e,w}^\prime$ & 0.0087 & $C_{ {s}}^{w}$ & $8.5\times 10^5$\\
		 $R_{n,s}^\prime$ & 0.0116 &&\\
		\hline
	\end{tabular}
\end{table}

\section{Control methodologies}
\subsection{Problem formulation}
\subsubsection{Centralized MPC formulation}
\begin{figure}[!htb]
	\centering
	\includegraphics[width=0.32\textwidth]{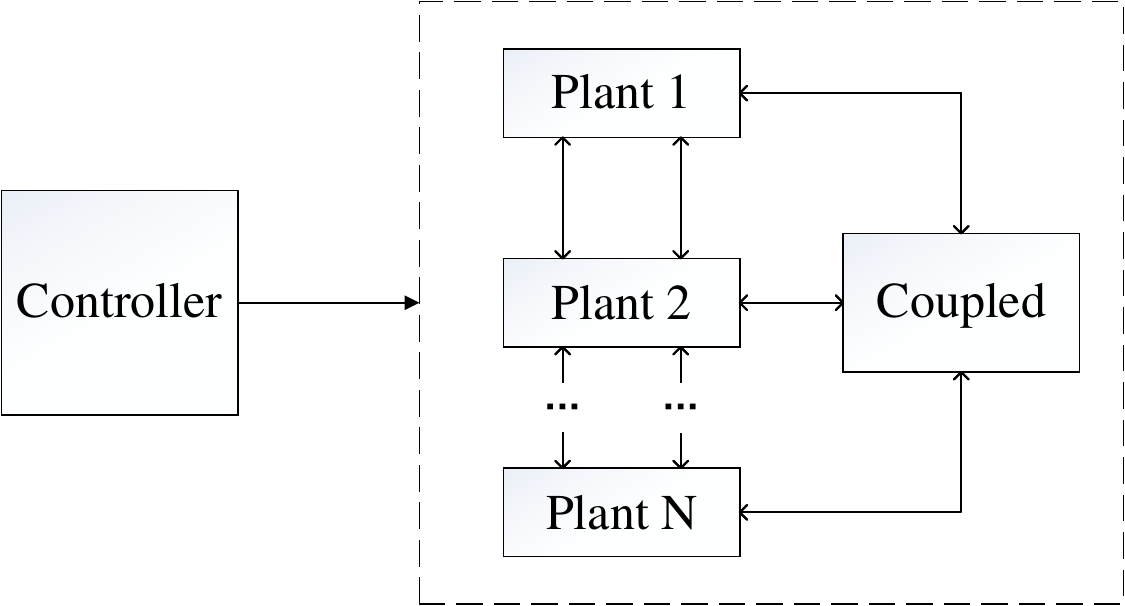}
	\caption{Centralized control framework.}
	\label{centralized1}
\end{figure}

The centralized control strategy is to design a master controller to coordinate the temperature of all zones. The centralized block diagram for $N$ zones is shown in Fig. \ref{centralized1}.

The centralized optimization objective needs to integrate all subsystems' objectives.
The optimization objective function consists of two components: one is to impose a soft constraint to make the room temperature comfortable; the other is to minimize energy consumption. 
The objective for centralized predictive control is designed as
\begin{equation}        
     J(k)=\sum\limits_{m \in \mathcal{M}}\Big[\theta _m (\alpha J_m^v + J_m^u
     )\Big],
\end{equation}
where $J^v_m$ and $J^u_m$ are the comfort cost and energy cost, respectively. $\mathcal{M}$ is the index set of all $N$ subsystems.
$\alpha$ is the weight, which can be set according to the user's preference (more energy efficient or more comfortable). 
$\theta_m$ is the weight that regulates the importance of energy levels between subsystems.
 The expressions for $J^v_m$ and $J^u_m$ are
\begin{equation*}
	\begin{array}{ll}
		&J_m^v=\sum\limits_{l=1}^P\left\|v_m(k+l)\right\|^2\cdot{\delta_m(k+l)},\\
        &J_m^u=\sum\limits_{l=1}^P\left\|u_m(k+l-1)\right\|^2\cdot{\lambda_m(k+l-1)},
	\end{array}
\end{equation*}
where $P$ is the total number of the time step. The effect of $v_m$ is to impose a soft constraint that allows the room temperature to be controlled to a suitable range.
$\delta_m$ takes the value of 1 when the building is occupied and 0 when the building is not occupied.
$u_m$ denotes the input power of the HVAC system. 
The electricity charge rate in Shenzhen is shown in Table \ref{fee}, and $\lambda_m$ is the weight for the time-varying electricity price.  

Furthermore, define variables that gather states, inputs and outputs over the prediction horizon: $\mathbf{x}=[x^\intercal(k+1),\:\cdots\:x^\intercal(k+P)]^\intercal$,
$\mathbf{v}_m=[v_m^\intercal(k+1),\:\cdots\:v_m^\intercal(k+P)]^\intercal$, $\mathbf{u}_m=[u_m^\intercal(k),\:\cdots\:u_m^\intercal(k+P-1)]^\intercal$, $\mathbf{y}_m=[y_m^\intercal(k+1),\:\cdots\:y_m^\intercal(k+P)]^\intercal$.
The sum of the inputs to the subsystem must be within the energy limit, we have
\begin{equation}
    \sum\limits_{m \in \mathcal{M}} \mathbf{u}_m \leq \mathbf{c}^{\max },
    \label{energy_lim}
\end{equation}
where $\mathbf{c}^{\max }$ denotes the total energy limit. 

The scheme for centralized predictive control is then designed as follows:
\begin{equation}
	\begin{array}{lll}
		&\min\limits_{\substack{\mathbf{u}_1,\cdots,\mathbf{u}_N\\\mathbf{v}_1,\cdots,\mathbf{v}_N}} & J(k)=\sum\limits_{m \in \mathcal{M}}\Big[\theta _m (\alpha J_m^v + J_m^u
     )\Big] \\
		&\text { s.t. } & x(k+1)=A x(k)+B u(k)+d(k) \\
		&& y(k)=C x(k) \\
		&&x(k)=x^* \\
		&&\sum\limits_{m \in \mathcal{M}} \mathbf{u}_m \leq \mathbf{c}^{\max } \\
		&&\text { For all } m \in \mathcal{M}: \\
		&&\mathbf{u}_m^{\min } \leq \mathbf{u}_m \leq \mathbf{u}_m^{\max } \\
		&&\mathbf{y}_m^{\min } \leq \mathbf{y}_m+\mathbf{v}_m \\
		&&\mathbf{y}_m^{\max } \geq \mathbf{y}_m-\mathbf{v}_m \\
		&&\mathbf{v}_m \geq 0  \\
	\end{array}
	\label{cent}
\end{equation}
where $x^*$ is the initial value of the state $x$ at the $k$-th moment.
$\mathbf{u}_m^{\min }$ and $\mathbf{u}_m^{\max}$ indicate the maximum and minimum input power. The RC model used in the centralized scheme is about the whole system with $N$ subsystems.

\begin{table}[htbp]			
	\centering
	\caption{Electricity charge rate in Shenzhen.}
	\label{fee} 
	\begin{tabular}{cc}
		\hline
		Time&Electricity price $\lambda$ (CNY/kWh)\\
		\hline
		0:00-8:00&0.3358\\
		8:00-14:00&0.6629\\
		14:00-17:00&1.0881\\
		17:00-19:00&0.6629\\
		19:00-22:00&1.0881\\
		22:00-24:00&0.6629\\
		\hline			
	\end{tabular}
\end{table}
\subsubsection{Decentralized MPC formulation}
\begin{figure}[!htb]
	\centering
	\includegraphics[width=0.32\textwidth]{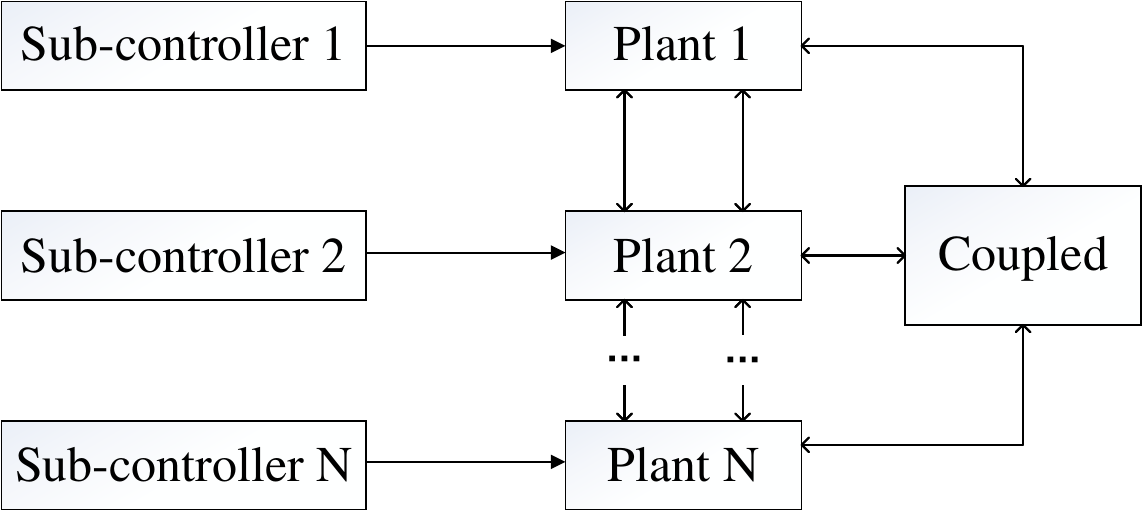}
	\caption{Decentralized control framework.}
	\label{decentralized1}
\end{figure}

In the decentralized scenario, each zone is configured with a sub-controller. Each sub-controller corresponds to a subsystem, and there are couplings between the subsystems, but the controllers are completely independent. The decentralized block diagram for $N$ zones is shown in Fig. \ref{decentralized1}.

The decentralized MPC objective is related only to the subsystem itself and is designed as
\begin{equation}
    J_m(k)=\alpha J_m^v + J_m^u.
\end{equation}
Since the sub-controllers are not clear to each other about their respective inputs. Constraint (\ref{energy_lim}) in the centralized scheme could be replaced with
\begin{equation}
    \mathbf{u}_m \leq \hat {\mathbf{c}}^{a },
    \label{energy_lim_de}
\end{equation}
where $\hat {\mathbf{c}}^{a }=1/N\cdot\mathbf{c}^{\max }$. Constraint (\ref{energy_lim_de}) implies that the energy allowances are divided equally among each subsystem in the decentralized scheme.

Then the decentralized MPC scheme for the $m$-th subsystem is as follows.
\begin{equation}
	\begin{array}{lll}
		&\min\limits_{\mathbf{u}_m,\mathbf{v}_m}&  J_m(k)=\alpha J_m^v + J_m^u
     \\
		&\text { s.t. } & x_m(k+1)=A_mx_m(k)+B_{m}u_m(k)\\
		&&\qquad\qquad\qquad+d_m(k) \\
		&& y_m(k)=C_mx_m(k) \\
		&&x_m(k)=x_m^* \\
		&& \mathbf{u}_m \leq \hat {\mathbf{c}}^{a } \\
		&&\mathbf{u}_m^{\min } \leq \mathbf{u}_m \leq \mathbf{u}_m^{\max } \\
		&&\mathbf{y}_m^{\min } \leq \mathbf{y}_m+\mathbf{v}_m \\
		&&\mathbf{y}_m^{\max } \geq \mathbf{y}_m-\mathbf{v}_m \\
		&&\mathbf{v}_m \geq 0  \\
	\end{array}
	\label{mpc2}
\end{equation}
It is noted that the RC model expression is only about the $m$-th subsystem.
Compared with centralized strategies, decentralized controllers deal with relatively simple optimization problems because the size of the objective function and constraints are reduced. However, the decentralized strategy is not flexible enough to handle the shared constraints and has insufficient scheduling capability for energy, which may result in insufficient supply for some rooms with high energy demand and energy redundancy for other rooms with low energy demand.

\subsubsection{Distributed MPC formulation}
\begin{figure}[!htb]
	\centering
	\includegraphics[width=0.44\textwidth]{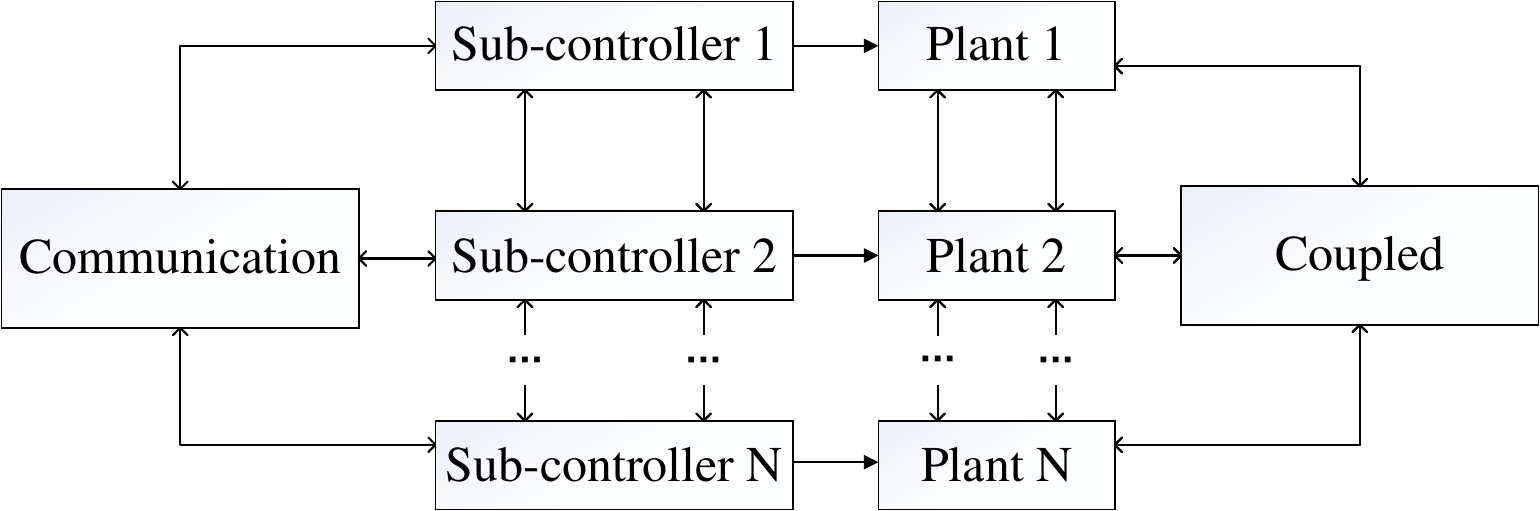}
	\caption{Distributed control framework.}
	\label{distributed1}
\end{figure}

With a distributed strategy, there is an exchange of information between a sub-controller and its neighbor sub-controllers. The controller makes better decisions by combining the information obtained from the neighbors. The distributed control framework is shown in Fig. \ref{distributed1}.

Compared with constraint (\ref{energy_lim}) and constraint (\ref{energy_lim_de}), the energy allowance in the distributed scheme is designed as 
\begin{equation}
    \mathbf{u}_m \leq \hat {\mathbf{c}}_{m }(k),
    \label{ddcc}
\end{equation}
where $\hat {\mathbf{c}}_{m }(k)$ gives an upper bound on the input $\mathbf{u}_m$. 

The distributed MPC scheme for subsystem $m$ is then as follows.
\begin{equation}
	\begin{array}{lll}
		&\min\limits_{\mathbf{u}_m,\mathbf{v}_m} & J_m(k)=\alpha J_m^v + J_m^u\\
		&\text { s.t. } &  
		x_m(k+1)=A_mx_m(k)+B_{m}u_m(k)\\
		&&\qquad\qquad\qquad+ d_m(k) \\
		&& y_m(k)=C_mx_m(k) \\
		&&x_m(k)=x_m^* \\
		&&\mathbf{u}_m \leq \hat {\mathbf{c}}_{m }(k) \\
		&&\mathbf{u}_m^{\min } \leq \mathbf{u}_m \leq \mathbf{u}_m^{\max } \\
		&&\mathbf{y}_m^{\min } \leq \mathbf{y}_m+\mathbf{v}_m \\
		&&\mathbf{y}_m^{\max } \geq \mathbf{y}_m-\mathbf{v}_m \\
		&&\mathbf{v}_m \geq 0  \\
	\end{array}
	\label{d1}
\end{equation}
As with the decentralized scheme, the distributed RC model expression involves only the $m$-th subsystem. The different between (\ref{mpc2}) and (\ref{d1}) is the constrain on $\mathbf{u}_m$.
The value of $\mathbf{c}_{m }(k)$ is the key to solving the problem (\ref{d1}), and the following section describes how to get a reasonable $\mathbf{c}_{m }(k)$.

\subsubsection{Differences among models}
We have given the centralized, decentralized and distributed formulations above as shown in (\ref{cent}), (\ref{mpc2}) and (\ref{d1}) respectively.
An RC model expression for the whole system is used for the prediction model of the centralized scheme, in contrast to the decentralized and distributed prediction models, which involve only the corresponding subsystem. Therefore, the size of the centralized MPC problem is larger compared with the other two schemes. The size of the centralized problem is related to the current size of the whole system, more precisely, when the system size increases, the size of the corresponding optimization problem also increases. The size of the decentralized and distributed MPC problems is usually fixed and does not increase with the size of the whole system, which implies that decentralized and distributed schemes are expected to be extended to larger-scale systems to obtain computationally tractable solutions.

Furthermore, the centralized MPC has an advantage in global optimality and overall performance because it has access to global information about the system. The decentralized MPC may have poorer control performance due to the lack of global information and information exchange. Distributed MPC is promising to balance global optimality and computational efficiency to achieve better control performance through coordination and information exchange between subsystems.

\subsection{Distributed algorithm}
\label{dist_section}
\subsubsection{One-to-one priority strategy}
\begin{figure}[!htb]
	\centering
	\includegraphics[width=0.4\textwidth]{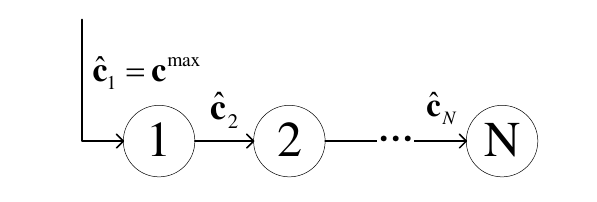}
	\caption{Information exchange between subsystems.}
	\label{distributed_energy}
\end{figure}

One-to-one priority means that each subsystem in the system corresponds to a different priority, i.e., $N$ subsystems correspond to $N$ priorities. The information exchange between subsystems in the one-to-one priority case is shown in Fig. \ref{distributed_energy}, where the subsystems are numbered from 1 according to the priority of energy distribution, and the top-numbered subsystem has a high priority for energy supply. 
Note that the information exchange shown in Fig. \ref{distributed_energy} is parallel.
For subsystem $m$, subsystem $m-1$ is defined as the upstream subsystem and subsystem $m+1$ is the downstream subsystem. $\hat{\mathbf{c}}_{m}$ satisfies $\hat{\mathbf{c}}_{m+1}=\hat{\mathbf{c}}_{m}-\mathbf{u}_m$.
For the first subsystem, the maximum energy limit is obtained, i.e. $\hat{\mathbf{c}}_{1}=\mathbf{c}^{max}$.

The sub-controllers of all subsystems can work in parallel. At moment $k$, subsystem $m$ gets $\hat{\mathbf{c}}_{m}(k)$ from the upstream subsystem $m-1$, then solves its own optimization problem and gets the solution $\mathbf{u}_m(k|k)$. Further, $\hat{\mathbf{c}}_{m+1}(k+1)$ is obtained by
\begin{equation}
 \hat{\mathbf{c}}_{m+1}(k+1)=\hat{\mathbf{c}}_{m}(k+1|k)-\mathbf{u}_m(k+1|k) ,
 \label{cm}
\end{equation}
 and is transmitted to the downstream subsystem $ m+1$, i.e.,
remove the first element of $\mathbf{u}_m(k|k)$ and move the remaining elements one step forward to get $\mathbf{u}_m(k+1|k)$. Note that the last two elements of $\mathbf{u}_m(k+1|k)$ are the same in order to keep the vector length consistent.
The quantities of $\mathbf{u}_m$ and $\mathbf{c}_m$ are shown in (\ref{11}).
See Algorithm \ref{alg1} for the one-to-one allocation algorithm. $K$ denotes the termination moment.
\begin{figure*}
	\begin{equation}
		\begin{array}{c}
			\mathbf{u}_m(k|k)=[u_m(k|k)\quad u_m(k+1|k)\quad\cdots\quad u_m(k+M-1|k)]^T\\
			\mathbf{u}_m(k+1|k)=[u_m(k+1|k)\quad u_m(k+2|k)\quad\cdots\quad u_m(k+M-1|k)\quad u_m(k+M-1|k)]^T\\
			\hat{\mathbf{c}}_{m}(k)=[\hat{c}_{m}(k|k)\quad \hat{c}_{m}(k+1|k)\quad\cdots\quad \hat{c}_{m}(k+M-1|k)]^T\\
			\hat{\mathbf{c}}_{m}(k+1|k)=[\hat{c}_{m}(k+1|k)\quad \hat{c}_{m}(k+2|k)\quad\cdots\quad \hat{c}_{m}(k+M-1|k)\quad \hat{c}_{m}(k+M-1|k)]^T\\
		\end{array}
		\label{11}
	\end{equation}
\end{figure*}

\begin{algorithm}[htb]
	\caption{One-to-one priority algorithm}
	\label{alg1}
	\begin{algorithmic}[1]
		\State Given the initial $\hat{\mathbf{c}}_1,\cdots,\hat{\mathbf{c}}_N$.
		\For {$k \in \{1,\cdots,K\}$}
		\ParFor {$m \in \mathcal{M}$}
		\State Get $\hat{\mathbf{c}}_m(k)$ from upstream subsystem $ m-1$.
		\State Solve (\ref{d1}) to obtain the solution $\mathbf{u}_m(k|k)$.
		\State Get $\mathbf{u}_m(k+1|k)$ and $\hat{\mathbf{c}}_{m}(k+1|k)$.
		\State  Get $\hat{\mathbf{c}}_{m+1}(k+1)=\hat{\mathbf{c}}_{m}(k+1|k)-\mathbf{u}_m(k+1|k)$ 
		\Statex \quad\quad\quad and transmit it to the downstream subsystem $ m+1$.
		\EndParFor
		\EndFor
	\end{algorithmic}
\end{algorithm}

It is noted that the above computational process uses the predicted solution of the current moment for the future moment, so the distributed approach allows all subsystems to work in parallel and the size of the optimization problem to be handled by each subsystem does not increase with the size of the system. 

\subsubsection{Multi-to-one priority strategy}
Multi-to-one priority means that multiple subsystems in a system can correspond to the same priority level. In large-scale building systems, there may be multiple zones corresponding to the same supply priority level, and the multiple-to-one priority strategy is suitable for this situation.

The information matrix $I_{nf}$ records the information about the remaining energy, and if the subsystem is divided into $N_o$ energy supply levels starting from level 1, then $I_{nf}$ can be expressed as,
\begin{equation*}
	I_{nf}=
	\left(\begin{array}{c}
		\frac{\mathbf{c}_{\max }}{N_{{pri}=1}} \\
		\frac{\mathbf{c}_{\max }-\sum\limits_{m:pri(m)=1} \mathbf{u_m}}{N_{{pri}=2}} \\
		\frac{\mathbf{c}_{\max }-\sum\limits_{m:pri(m)=1} \mathbf{u_m}-\sum\limits_{m:pri(m)=2} \mathbf{u_m}}{N_{pri=3}}\\
		\vdots\\
		\frac{\mathbf{c}_{\max }-\sum\limits_{m:pri(m)<s} \mathbf{u_m}}{N_{pri=s}}\\
		\vdots\\
		\frac{\mathbf{c}_{\max }-\sum\limits_{m:pri(m)<N_o} \mathbf{u_m}}{N_{pri=N_o}}
	\end{array}\right),
	\label{INF}
\end{equation*}
where $pri(m)$ is the index function of the priority, $\sum\limits_{m:pri(m)<s} \mathbf{u_m}$ denotes the sum of the solutions for subsystems with priority less than $s$, ${N_{pri=s}}=|\{m|pri(m)=s\}|$ indicates the number of subsystems with priority $s$. $I_{nf}$ has $N_o$ rows and $M$ columns, and the last two columns have the same elements. For subsystem $m$, if its priority $pri(m) = s$, then $\hat {\mathbf{c}}_{m }$ records its energy residual information, $\hat {\mathbf{c}}_{m }=I_{nf}(s)$, $1\leq s\leq N_o$, and is the element of the $s$-th row of $I_{nf}$.

\begin{figure}[!htb]
	\centering
	\includegraphics[width=0.35\textwidth]{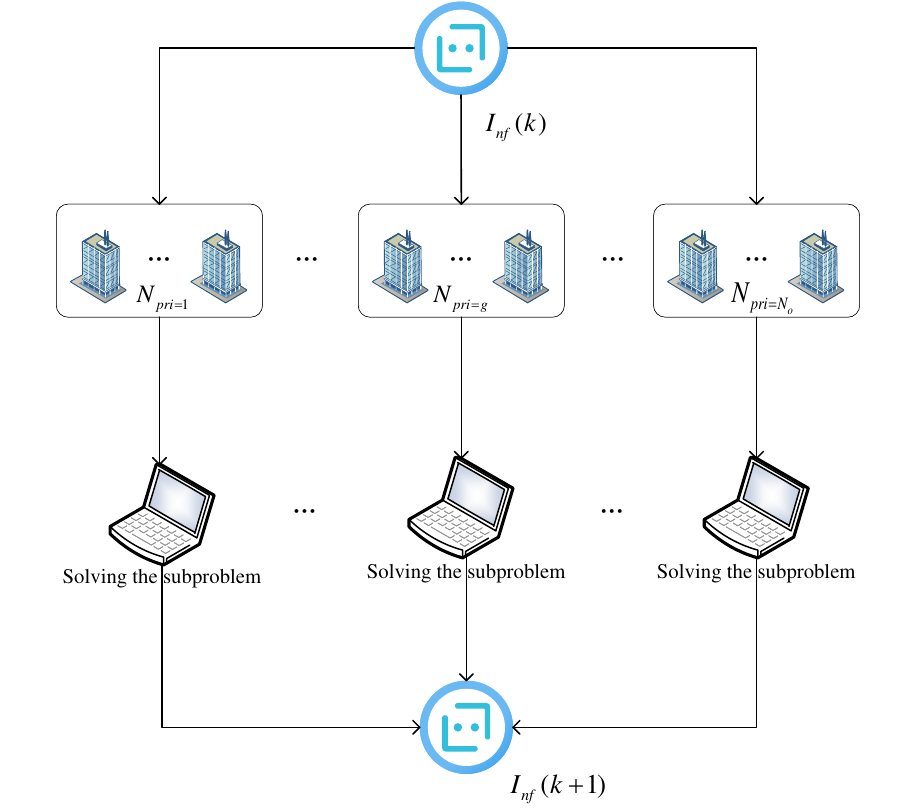}
	\caption{Information exchange for multi-to-one priority strategy.}
	\label{multi-to-one}
\end{figure}

As Fig. \ref{multi-to-one} shown, each subsystem visits $I_{nf}(k)$ at the $k$-th moment, see (\ref{INFk}).
\begin{figure*}
	\begin{equation}
		I_{nf}(k)=
		\left(\begin{array}{cccc}
			\frac{c_{\max }}{N_{{pri}=1}}&\frac{c_{\max }}{N_{{pri}=1}}&\cdots&\frac{c_{\max }}{N_{{pri}=1}} \\
			\frac{c_{\max }-\sum\limits_{m:pri(m)=1} u_m(k|k-1)}{N_{{pri}=2}}&\frac{c_{\max }-\sum\limits_{m:pri(m)=1} u_m(k+1|k-1)}{N_{{pri}=2}}&\cdots &\frac{c_{\max }-\sum\limits_{m:pri(m)=1} u_m(k+M-2|k-1)}{N_{{pri}=2}} \\
			\vdots & \vdots & \vdots & \vdots \\
			\frac{c_{\max }-\sum\limits_{m:pri(m)<s} u_m(k|k-1)}{N_{pri=s}}&\frac{c_{\max }-\sum\limits_{m:pri(m)<s} u_m(k+1|k-1)}{N_{pri=s}} & \cdots & \frac{c_{\max }-\sum\limits_{m:pri(m)<s} u_m(k+M-2|k-1)}{N_{pri=s}}\\
			\vdots & \vdots & \vdots & \vdots \\
			\frac{c_{\max }-\sum\limits_{m:pri(m)<N_o} u_m(k|k-1)}{N_{pri=N_o}}&\frac{c_{\max }-\sum\limits_{m:pri(m)<N_o} u_m(k+1|k-1)}{N_{pri=N_o}} & \cdots & \frac{c_{\max }-\sum\limits_{m:pri(m)<N_o} u_m(k+M-2|k-1)}{N_{pri=N_o}}
		\end{array}\right).
		\label{INFk}
	\end{equation}
\end{figure*}
The corresponding $\hat{\mathbf{c}}$ is obtained, and then the optimization problem is solved and $I_{nf}$ is updated using the new solution to obtain $I_{nf}(k+1)$, see (\ref{INFk1}).
\begin{figure*}
	\begin{equation}
		I_{nf}(k+1)=
		\left(\begin{array}{cccc}
			\frac{c_{\max }}{N_{{pri}=1}}&\frac{c_{\max }}{N_{{pri}=1}}&\cdots&\frac{c_{\max }}{N_{{pri}=1}} \\
			\frac{c_{\max }-\sum\limits_{m:pri(m)=1} u_m(k+1|k)}{N_{{pri}=2}}&\frac{c_{\max }-\sum\limits_{m:pri(m)=1} u_m(k+2|k)}{N_{{pri}=2}}&\cdots &\frac{c_{\max }-\sum\limits_{m:pri(m)=1} u_m(k+M-1|k)}{N_{{pri}=2}} \\
			\vdots & \vdots & \vdots & \vdots \\
			\frac{c_{\max }-\sum\limits_{m:pri(m)<s} u_m(k+1|k)}{N_{pri=s}}&\frac{c_{\max }-\sum\limits_{m:pri(m)<s} u_m(k+2|k)}{N_{pri=s}} & \cdots & \frac{c_{\max }-\sum\limits_{m:pri(m)<s} u_m(k+M-1|k)}{N_{pri=s}}\\
			\vdots & \vdots & \vdots & \vdots \\
			\frac{c_{\max }-\sum\limits_{m:pri(m)<N_o} u_m(k+1|k)}{N_{pri=N_o}}&\frac{c_{\max }-\sum\limits_{m:pri(m)<N_o} u_m(k+2|k)}{N_{pri=N_o}} & \cdots & \frac{c_{\max }-\sum\limits_{m:pri(m)<N_o} u_m(k+M-1|k)}{N_{pri=N_o}}
		\end{array}\right).
		\label{INFk1}
	\end{equation}
\end{figure*}
When ${N_{pri=1}}={N_{pri=2}}=\cdots={N_{pri=N_o}}=1$, it degenerates to the case of one-to-one priority. It is noted that this distributed algorithm makes full use of the whole set of solutions in the prediction horizon, where the first element is applied to the plant and the others are used to update the information matrix $I_{nf}$. See Algorithm \ref{alg2} for the multi-to-one allocation algorithm.

\begin{algorithm}[htb]
	\caption{Multi-to-One priority algorithm}
	\label{alg2}
	\begin{algorithmic}[1]
		\State Given the initial $I_{nf}$.
		\For {$k \in \{1,\cdots,K\}$}
		\State Get $I_{nf}(k)$.
		\ParFor {$m \in \mathcal{M}$}
		\State Get $\hat{\mathbf{c}}_m(k)$ from $I_{nf}(k)$ according to the priority 
		
		\Statex \quad\quad\quad of $m$.
	    \State Solve (\ref{d1}) to obtain the solution $\mathbf{u}_m(k|k)$.
	    \State Get $\mathbf{u}_m(k+1|k)$.
	    \EndParFor
		\State Update to $I_{nf}(k+1)$.
		\EndFor
	\end{algorithmic}
\end{algorithm}

\subsection{Strategic Analysis}
\subsubsection{Performance of subsystems with the highest priority}

Compared with the centralized scheme, in addition to reducing the problem size, our distributed scheme is similar to giving subsystems with higher priority a considerable weight in energy allocation, since the energy amount of subsystems with lower priority depends on the estimated residual energy of higher priority systems.
We give a lemma and its proof to explain this statement.

\newtheorem{lemma}{Lemma}

\begin{lemma}
	Denote$\{\mathbf{u}_{1c}^*,\mathbf{v}_{1c}^*,\cdots,\mathbf{u}_{nc}^*,\mathbf{v}_{nc}^*\}$ as the optimal solution for the centralized optimization problem (\ref{cent}), $\mathbf{u}_{1dc}^*,\mathbf{v}_{1dc}^*$ as the optimal solution for the decentralized optimization problem (\ref{mpc2}), and $\mathbf{u}_{1d}^*,\mathbf{v}_{1d}^*$ as the optimal solution for the distributed optimization problem (\ref{d1}), we have $J_1(\mathbf{u}_{1d}^*,\mathbf{v}_{1d}^*)\leq J_1(\mathbf{u}_{1dc}^*,\mathbf{v}_{1dc}^*)$ and $J_1(\mathbf{u}_{1d}^*,\mathbf{v}_{1d}^*)\leq J_1(\mathbf{u}_{1c}^*,\mathbf{v}_{1c}^*)$.i.e., the proposed distributed strategy perform better than or equal to the decentralized and centralized strategies for optimizing $J_1$.
	\label{lemma}
\end{lemma}

\newtheorem{proof}{Proof}
\begin{proof}
	Firstly, compare the optimization of the distributed and decentralized algorithms for $J_1$.
	Consider (\ref{mpc2}) with $\hat {\mathbf{c}}^{a }=1/N\cdot\mathbf{c}^{\max }$ and (\ref{d1}) with $\hat{\mathbf{c}}_{1}=\mathbf{c}^{max}$, we know that the feasible domain of the distributed problem (\ref{d1}) contains the feasible domain of the decentralized problem (\ref{mpc2}).
	Therefore, $J_1(\mathbf{u}_{1d}^*,\mathbf{v}_{1d}^*)\leq J_1(\mathbf{u}_{1dc}^*,\mathbf{v}_{1dc}^*)$.
	
	Then, compare the optimization of the distributed and centralized algorithms for $J_1$.
	The objective function of the centralized strategy can be expressed as
\begin{equation*}
	\begin{array}{ll}
	J=&J_1\left(\mathbf{u}_1,\mathbf{v}_1\right)+\frac{\theta_2}{\theta_1} J_2\left(\mathbf{u}_2,\mathbf{v}_2\right)+\frac{\theta_3}{\theta_1} J_3\left(\mathbf{u}_3,\mathbf{v}_3\right)\\
	&+\cdots+\frac{\theta_n}{\theta_1} J_n\left(\mathbf{u}_N,\mathbf{v}_N\right) ,
	\end{array}
\end{equation*}
	where $\mathbf{u}_1+\mathbf{u}_2+\cdots+\mathbf{u}_N \leq \mathbf{c}_{\max }$ and
	$ \theta_1 \geq \theta_2 \geq \cdots \geq \theta_N>0$.
	Since $J_m \geq 0$, optimizing $J_1$ alone is better or equal to optimizing $J$ for $J_1$. So the following optimization problem (\ref{dd1}) is better or equal to the centralized strategy for optimizing  $J_1$.
	
\begin{equation}
	\begin{array}{lll}
		&\min\limits_{\mathbf{u}_1,\mathbf{v}_1} & J_1(\mathbf{u}_1,\mathbf{v}_1)\\
	    &\text { s.t. } &  
		X_1(k+1)=A_1X_1(k)+B_{u1}u_1(k)\\
		&&\qquad\qquad\qquad+B_{d1} d_1(k) \\
		&& y_1(k)=C_1X_1(k) \\
		&&X_1(k)=X_1^* \\
		&&  
		\mathbf{u}_1+\mathbf{u}_2+\cdots+\mathbf{u}_n \leq \mathbf{c}^{\max }\\
		&&\mathbf{u}_1^{\min } \leq \mathbf{u}_1 \leq \mathbf{u}_1^{\max } \\
		&&\mathbf{y}_1^{\min } \leq \mathbf{y}_1+\mathbf{v}_1 \\
		&&\mathbf{y}_1^{\max } \geq \mathbf{y}_1-\mathbf{v}_1 \\
		&&\mathbf{v}_1 \geq 0  \\
	\end{array}
	\label{dd1}
\end{equation}

Denote$\{\mathbf{u}_{1}^*,\mathbf{v}_{1}^*\}$ as the optimal solution for the optimization problem (\ref{dd1}), we have 
$J_1(\mathbf{u}_{1}^*,\mathbf{v}_{1}^*)\leq J_1(\mathbf{u}_{1c}^*,\mathbf{v}_{1c}^*)$.
	Consider (\ref{d1}) with $\hat{\mathbf{c}}_{1}=\mathbf{c}^{max}$, we know that the feasible domain of the distributed problem (\ref{d1}) when $m=1$ contains the feasible domain of the problem (\ref{dd1}), so we have
	$J_1(\mathbf{u}_{1d}^*,\mathbf{v}_{1d}^*)\leq J_1(\mathbf{u}_{1}^*,\mathbf{v}_{1}^*)$.
	Therefore, $J_1(\mathbf{u}_{1d}^*,\mathbf{v}_{1d}^*)\leq J_1(\mathbf{u}_{1c}^*,\mathbf{v}_{1c}^*)$.
	In summary, we get Lemma \ref{lemma}.
\end{proof}

\subsubsection{Performance analysis}
\label{3.4}

This part gives the performance analysis in terms of the Pareto fronts of the three strategies.
A three-priority example is given to show the performance analysis.

\begin{figure*}[htbp]
	\centering    
	\subfigure[Pareto front for the first priority.] 
	{
		\begin{minipage}[t]{0.3\linewidth}
			\centering          
			\includegraphics[width=1\textwidth]{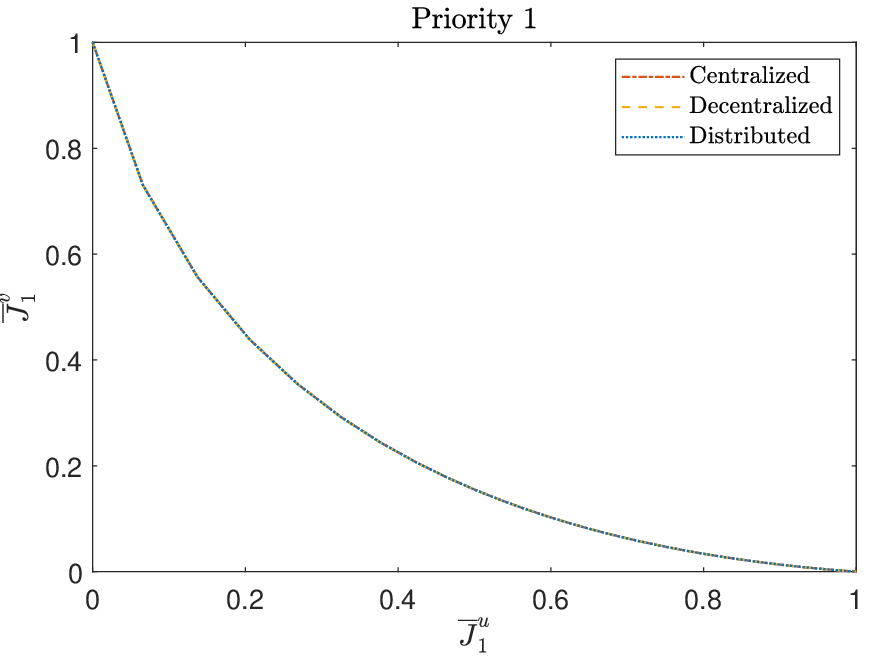}   
		\end{minipage}
		\label{priority 1}
	}
	\subfigure[Pareto front for second priority.] 
	{
		\begin{minipage}[t]{0.3\linewidth}
			\centering      
			\includegraphics[width=1\textwidth]{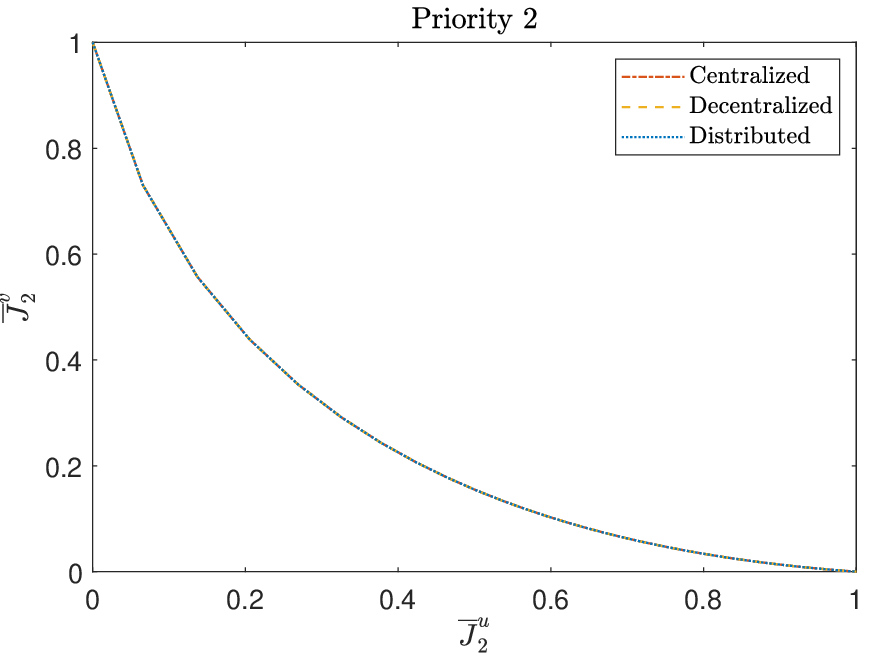}   
		\end{minipage}
		\label{priority 2}
	}
	\subfigure[Pareto front for third priority.] 
	{
	\begin{minipage}[t]{0.3\linewidth}
		\centering      
		\includegraphics[width=1\textwidth]{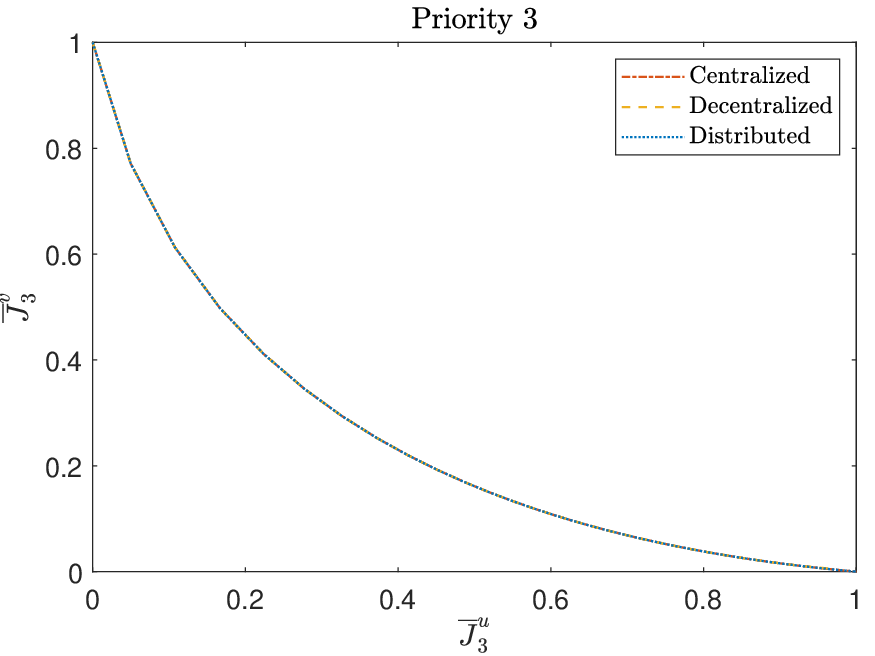}   
	\end{minipage}
	\label{priority 3}
	}
	\caption{Pareto fronts for three schemes (decoupled).} 
\label{fig5}  
\end{figure*}

\begin{figure*}[htbp]
	\centering    
	\subfigure[Pareto front for the first priority.] 
	{
		\begin{minipage}[t]{0.3\linewidth}
			\centering          
			\includegraphics[width=1\textwidth]{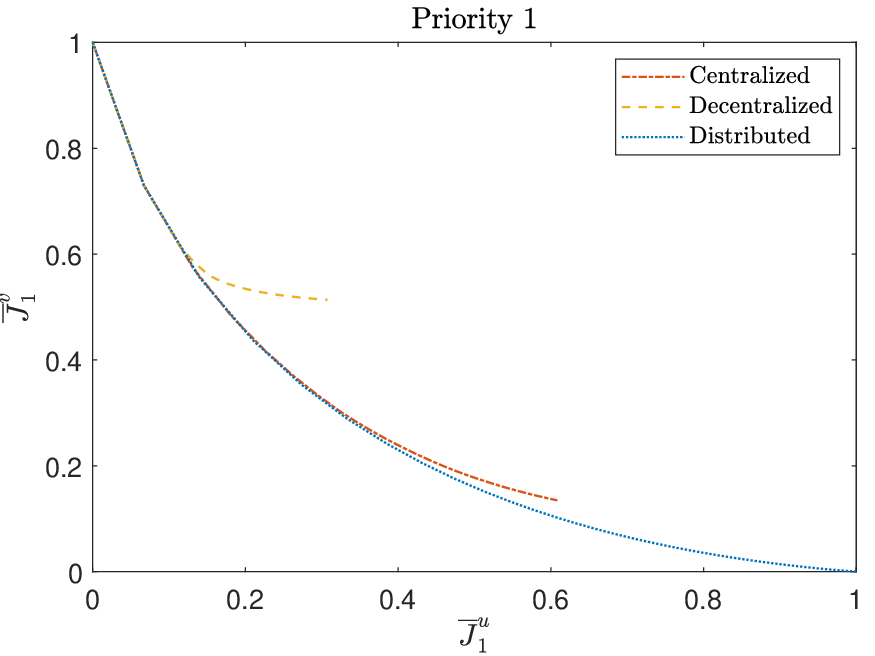}   
		\end{minipage}
		\label{priority 1lack0.50.2}
	}
	\subfigure[Pareto front for second priority.] 
	{
		\begin{minipage}[t]{0.3\linewidth}
			\centering      
			\includegraphics[width=1\textwidth]{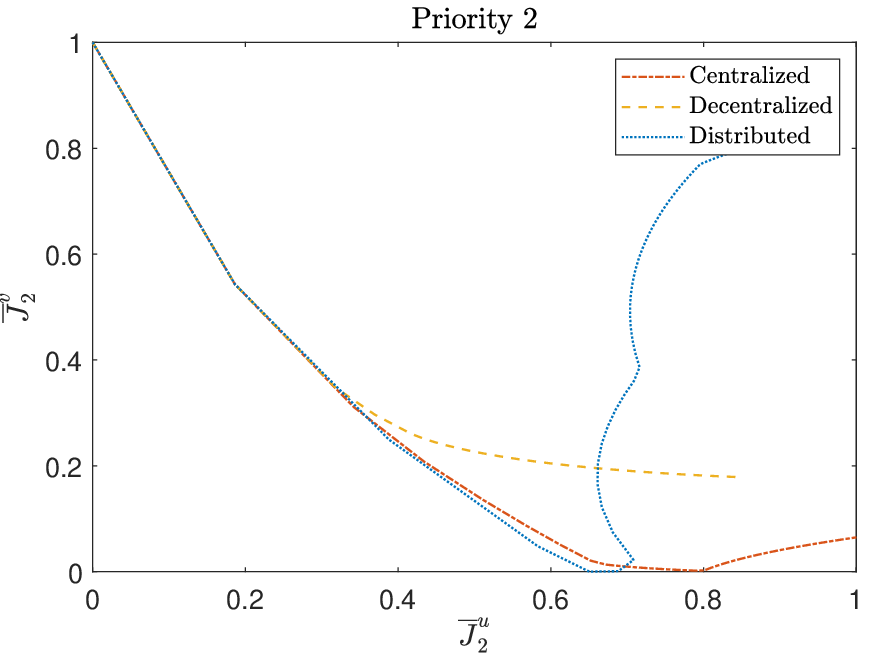}   
		\end{minipage}
		\label{priority 2lack0.50.2}
	}
	\subfigure[Pareto front for third priority.] 
	{
		\begin{minipage}[t]{0.3\linewidth}
			\centering      
			\includegraphics[width=1\textwidth]{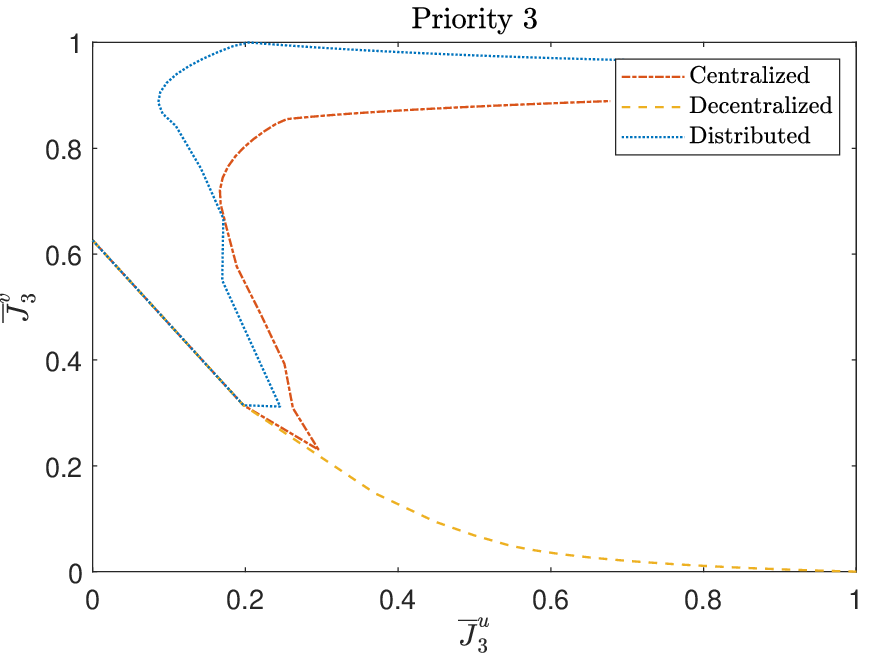}   
		\end{minipage}
		\label{priority 3lack0.50.2}
	}
	\caption{Pareto fronts for three schemes (coupled, $\theta_1=1, \theta_2=0.5, \theta_3=0.2$).} 
	\label{fig5lack0.50.1}  
\end{figure*}

\begin{figure*}[htbp]
	\centering    
	\subfigure[Pareto front for first priority.] 
	{
		\begin{minipage}[t]{0.3\linewidth}
			\centering          
			\includegraphics[width=1\textwidth]{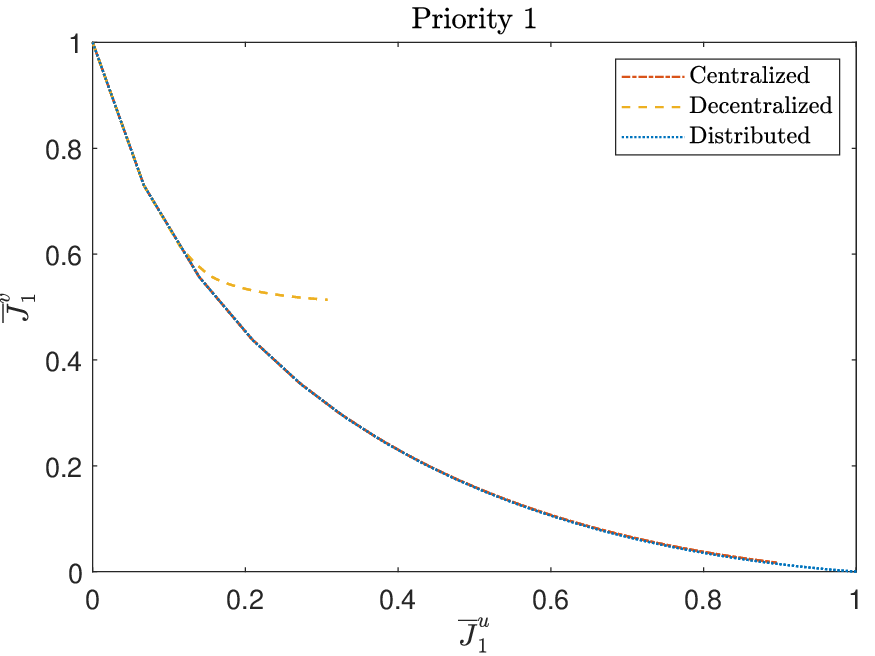}   
		\end{minipage}
		\label{priority 1lack}
	}
	\subfigure[Pareto front for second priority.] 
	{
		\begin{minipage}[t]{0.3\linewidth}
			\centering      
			\includegraphics[width=1\textwidth]{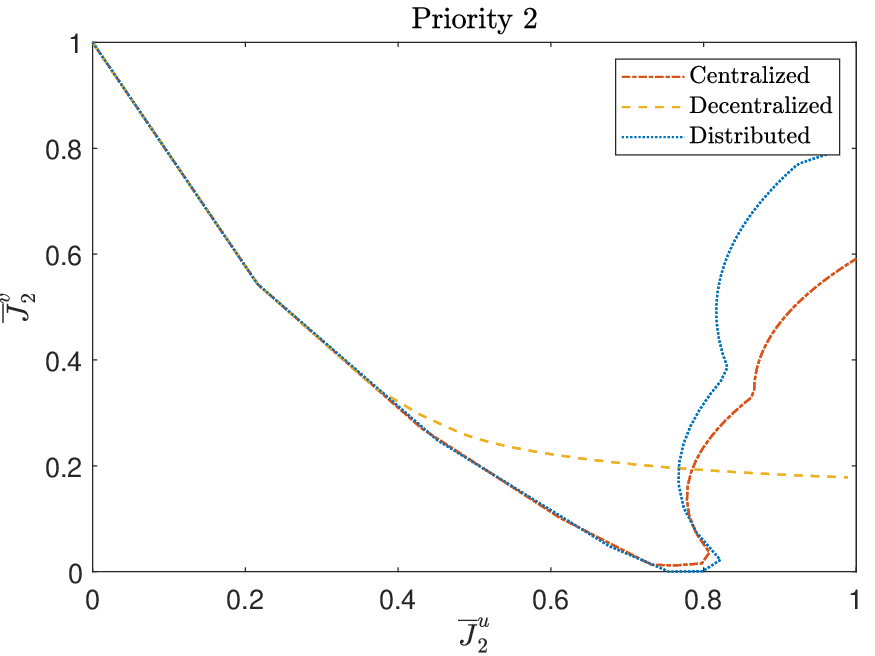}   
		\end{minipage}
		\label{priority 2lack}
	}
	\subfigure[Pareto front for third priority.] 
	{
		\begin{minipage}[t]{0.3\linewidth}
			\centering      
			\includegraphics[width=1\textwidth]{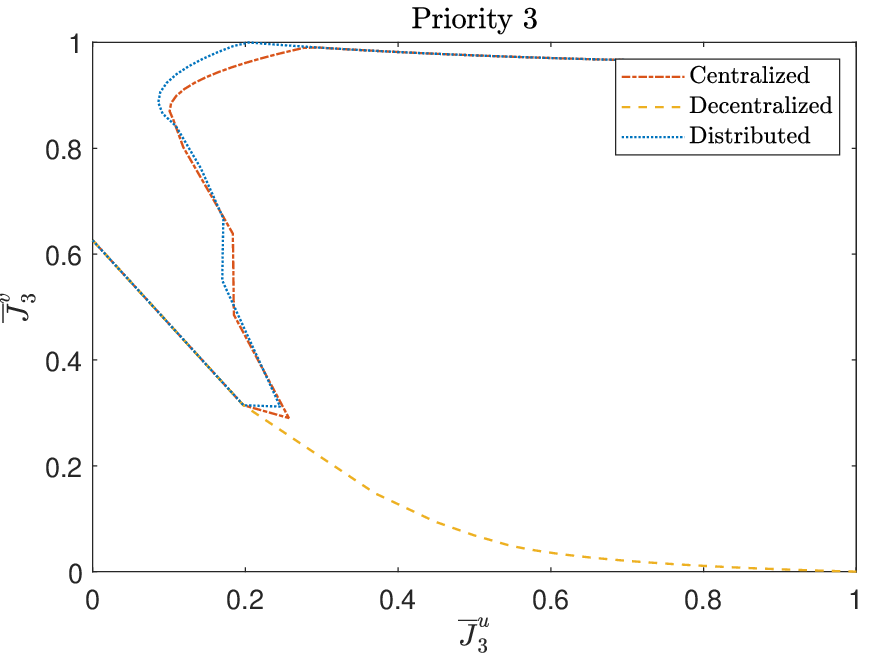}   
		\end{minipage}
		\label{priority 3lack}
	}
	\caption{Pareto fronts for three schemes (coupled, $\theta_1=1, \theta_2=0.1, \theta_3=0.01$).} 
	\label{fig5lack}  
\end{figure*}

\newtheorem{example}{Example}
\begin{example}
In this example, we analyze problems (\ref{cent}), (\ref{mpc2}) and (\ref{d1}) from a multi-objective optimization perspective. There are three subsystems in this example, each corresponding to a different priority. 

For the centralized scheme, only one optimization problem is constructed to correspond to the whole system, and the optimization objective is
\begin{equation}
    J(k)=\sum\limits_{m=1}^3\Big[\theta _m (\alpha J_m^v + J_m^u)\Big].
\end{equation}
There are six optimization objectives to be minimized, including $J_1^v$, $J_1^u$, $J_2^v$, $J_2^u$, $J_3^v$, $J_3^u$.

For the decentralized and distributed scheme, the optimization problem constructed for each subsystem is only about itself, and the objective is
\begin{equation}
    J_m(k)=\alpha J_m^v + J_m^u, \quad m\in\{{1,2,3}\}.
\end{equation}
Three optimization problems are involved in both decentralized and distributed schemes. For the first priority subsystem, the objective is to minimize $J_1^v$, $J_1^u$. Similarly, for the second and third priority subsystems, the objective is to minimize $J_2^v$, $J_2^u$ and $J_3^v$, $J_3^u$ respectively.

In the following, we analyze the optimization of each priority by varying the weight $\alpha$, i.e., the relations between $J_1^v$ and $J_1^u$, $J_2^v$ and $J_2^u$, $J_3^v$ and $J_3^u$, respectively.
Since the centralized scheme has six optimization objectives and our analysis deals with two of them at a turn, the analysis for the centralized one actually takes a two-dimensional projection of its Pareto front. For minimization problems, Pareto fronts close to the origin represent better solutions.

The optimization problem (\ref{cent}) is actually decoupled when $\mathbf{c}^{\max }$ is taken to be a very large value, i.e. $\mathbf{c}^{\max }=[2500,2500,\cdots,2500]$, and then this energy-limited constraint is always satisfied. By changing the weight $\alpha$, Fig. \ref{fig5} shows that in this case, the Pareto fronts of the three schemes overlap. 
From another point of view, if $\mathbf{c}^{\max }$ is large enough, there is practically no energy distribution problem for all three schemes since there is always enough energy to satisfy the comfort requirements.
 
Modifying the energy limit $\mathbf{c}^{\max }$ to be a smaller value, i.e. $\mathbf{c}^{\max }=[800,800,\cdots,800]$. There are energy couplings in subsystems because the existing energy can not satisfy the energy demand of all the rooms. The Pareto fronts are then made for simulation verification as shown in Fig. \ref{fig5lack0.50.1} ($\theta_1=1, \theta_2=0.5, \theta_3=0.2$) and Fig. \ref{fig5lack} ($\theta_1=1, \theta_2=0.1, \theta_3=0.01$). Due to the insufficient use of information, energy is equally allocated to the subsystems in the distributed scheme, resulting in poor optimization performance. The distributed scheme performs best for the optimization of the first priority subsystem, which supports Lemma \ref{lemma}.
As the values of $\theta_2, \theta_3$ decrease, the centralized Pareto front of the second and third priority subsystems approaches the distributed Pareto front.
This is due to the mechanisms of the two schemes. The mechanism of the distributed scheme is that the second priority subsystem is able to use all the estimated residual energy of the first priority subsystem, which means that the optimization of the second priority subsystem is carried out without taking into account the optimization of the subsystems with a lower priority than it, i.e., if the energy limit is very low, there is a possibility that the second priority subsystem will consume all the estimated residual energy to satisfy its optimization task. 
The centralized mechanism uses weights to distribute the energy of each priority. Therefore, when $\theta_2$ is small and $ \theta_3$ is much smaller than $\theta_2$, the performance of the distributed and centralized subsystems at second and third priority becomes close.

A fact is that as $\alpha$ becomes larger, the importance of the comfort cost $J_m^v$ increases and the demand for $\mathbf{u}_m$ increases, which means that the optimization problem is solved to get a larger $\mathbf{u}_m$ in this situation. It can be observed from Fig. \ref{priority 2lack0.50.2}, \ref{priority 3lack0.50.2}, \ref{priority 2lack} and \ref{priority 3lack} that as $\alpha$ becomes larger (along the x-axis), the distributed Pareto front becomes non-convex in the second and third priority cases.
This is because as $\alpha$ becomes larger, for the first priority $\mathbf{u}_1$ becomes larger, and from (\ref{cm}) we get $\hat{\mathbf{c}}_{2}$ becomes smaller and the constraints on (\ref{ddcc}) tighten. Similarly, for the third priority distributed optimization problem there is a tightening of the constraints (\ref{ddcc}), which results in a non-convex situation.
\end{example}

\section{Case study}
\subsection{Comfort index}
The comfort index is used to measure the degree of user satisfaction with the indoor temperature and is expressed by introducing a temperature deviation $e(t)$.
$e(t)$ is defined as 0 when the indoor temperature is in the specified temperature range. If not, $e(t)$ is the positive distance between the indoor temperature and the comfort range.
\begin{equation*}
	e(t)=\left\{\begin{array}{l}
		y_{\min }(t)-y(t), \quad \text { if } \:y(t)<y_{\min }(t) \\
		0, \,\,\,\quad\quad\quad\quad\quad\quad\text { if } \:y_{\min }(t)<y(t)<y_{\max }(t) \\
		y(t)-y_{\max }(t), \quad \text { if } \:y(t)>y_{\max }(t).
	\end{array}\right.
\end{equation*}
The comfort index $I_{ci}$ \cite{morocsan2010building} is defined as,
\begin{equation*}
	I_{ci}=\frac{1}{K}\sum_{\text {Occupation }}|e_i(t)|.
\end{equation*}
The subscript $i$ denotes the $i$-th priority. $K$ denotes the step. $I_{ci}$ represents the average deviation of the $i$-th priority subsystem beyond the comfort temperature range, the smaller the value the more comfortable the user feels.
For overall evaluation of comfort,
 $I_{c0}$ is defined as,

\begin{equation*}
	I_{c0}=\sqrt{\theta_1I_{c1}^2+\theta_2I_{c2}^2+\dots +\theta_nI_{cn}^2}.
\end{equation*}
The definition of $I_{c0}$ depends on the formulation of ${v}_m$ in the optimization object of the problem (\ref{cent}), where ${v}_m$ restricts the deviation between the indoor temperature and the comfort range.
Since $\theta$ represents the priority weight of each subsystem and $I_{ci}$ denotes the comfort level of the $i$-th subsystem, $I_{c0}$ can reflect the overall comfort level.
 A smaller value of $I_{c0}$ means that the building is more comfortable at an overall level.
 
\subsection{Small-scale scenario}
\begin{figure}[htbp]
	\centering
	\includegraphics[width=0.45\textwidth]{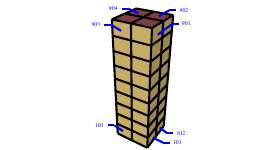} 
	\caption{ The building considered in the small-scale scenario.} 
	\label{Untitled} 
\end{figure}

In this scenario, a building model is built in EnergyPlus to replace the real building, which is $3.5m$ high and divided into three zones. The modeled building is shown in Fig.\ref{Untitled}, where Zone\:1 takes the highest priority for energy supply, Zone\:2 follows, and Zone\:3 takes the lowest priority. Considering the electricity prices shown in Table \ref{fee}, the comfortable temperature ranges for each zone at different moments are specified as shown in Table \ref{temp}.

\begin{table}[htbp]
	\centering
	\caption{Comfortable temperature range in the small-scale scenario}
	\label{temp}
	\small 
	\begin{tabular}{cccc}
		\hline 
		Time&Zone\:1&Zone\:2&Zone\:3\\
		\hline  
		0:00-10:00&no limit&no limit&no limit\\
		10:00-14:00&22-24$^{\circ}$C&22-24.5$^{\circ}$C&22-25$^{\circ}$C\\
		14:00-17:00&22-25$^{\circ}$C&22-25.5$^{\circ}$C&22-26$^{\circ}$C\\
		17:00-19:00&22-24$^{\circ}$C&22-24.5$^{\circ}$C&22-25$^{\circ}$C\\
		19:00-20:00&22-25$^{\circ}$C&22-25.5$^{\circ}$C&22-26$^{\circ}$C\\
		20:00-24:00&no limit&no limit&no limit\\
		\hline
	\end{tabular}
\end{table}

Each zone has the same area and is equipped with an ideal variable air volume terminal device, where the supply air flow can be changed from 0 to maximum value to meet the heating or cooling load of the zone. The main parameters of the building are shown in Table \ref{tab1}. 
\begin{table}[!htb]
	\centering
	\caption{Major building parameters}
	\label{tab1}
	\begin{tabular}{cc}
		\hline
		Building parameters&Preferences \\
		\hline
		Floor area&36m$^2$ \\
		Window to wall ratio&0.17 \\
		Occupant&1 occupant/12m$^2$ \\
		Lighting&0.75watts/m$^2$ \\
		Equipment&0.4watts/m$^2$ \\
		Occupied hours&10:00am-20:00pm \\
		\hline
	\end{tabular}
\end{table}
\subsection{Large-scale scenario}
\begin{figure}[htbp]
	\centering
	\includegraphics[width=0.32\textwidth]{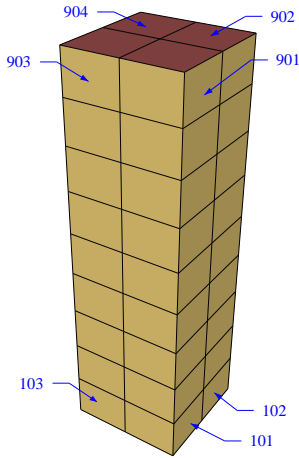} 
	\caption{ The building considered in the large-scale scenario.} 
	\label{Untitled2} 
\end{figure}

The considered large-scale scenario contains 36 zones, and the building model for the simulation experiment is shown in Fig. \ref{Untitled2}. The building has 9 floors, each of which has the same layout, and the $x$-th floor consists of four zones, $x$01, $x$02, $x$03, and $x$04, where $x$ is an integer between 1 and 9. The priority classification is shown in Table \ref{priority}, and the suitable temperature interval of each zone is shown in Table \ref{temp9}. The computing device used for the simulation experiment is an Intel(R) Xeon(R) Gold 6248R CPU @ 3.00GHz.

\begin{table}[htbp]
	\centering
	\caption{Priority of zones}
	\label{priority}
	\small 
	\begin{tabular}{ccc}
		\hline 
		first Priority&second Priority&third Priority\\
		\hline
		101, 102, & 301, 302, & \\
		103, 104, & 303, 304, &  The other zones \\
		(All zones on floor 1) & (All zones on floor 3) &\\
		\hline				
	\end{tabular}
\end{table}

\begin{table}[htbp]
	\centering
	\caption{Comfortable temperature range in the large-scale scenario}
	\label{temp9}
	\small 
	\begin{tabular}{ccccc}
		\hline
		Time&x01&x02&x03&x04\\
		\hline
		0:00-10:00&no limit&no limit&no limit&no limit\\
		10:00-14:00&22-24$^{\circ}$C&22-24.5$^{\circ}$C&22-25$^{\circ}$C&22-25.5$^{\circ}$C\\
		14:00-17:00&22-25$^{\circ}$C&22-25.5$^{\circ}$C&22-26$^{\circ}$C&22-26.5$^{\circ}$C\\
		17:00-19:00&22-24$^{\circ}$C&22-24.5$^{\circ}$C&22-25$^{\circ}$C&22-25.5$^{\circ}$C\\
		19:00-20:00&22-25$^{\circ}$C&22-25.5$^{\circ}$C&22-26$^{\circ}$C&22-26.5$^{\circ}$C\\
		20:00-24:00&no limit&no limit&no limit&no limit\\
		\hline				
	\end{tabular}
\end{table}

\section{Results}

\subsection{Test results in the small-scale scenario}

\begin{figure*}[htbp]
	\centering    
	\subfigure[Centralized strategy.] 
	{
		\begin{minipage}[t]{0.9\linewidth}
			\centering          
			\includegraphics[width=1\textwidth]{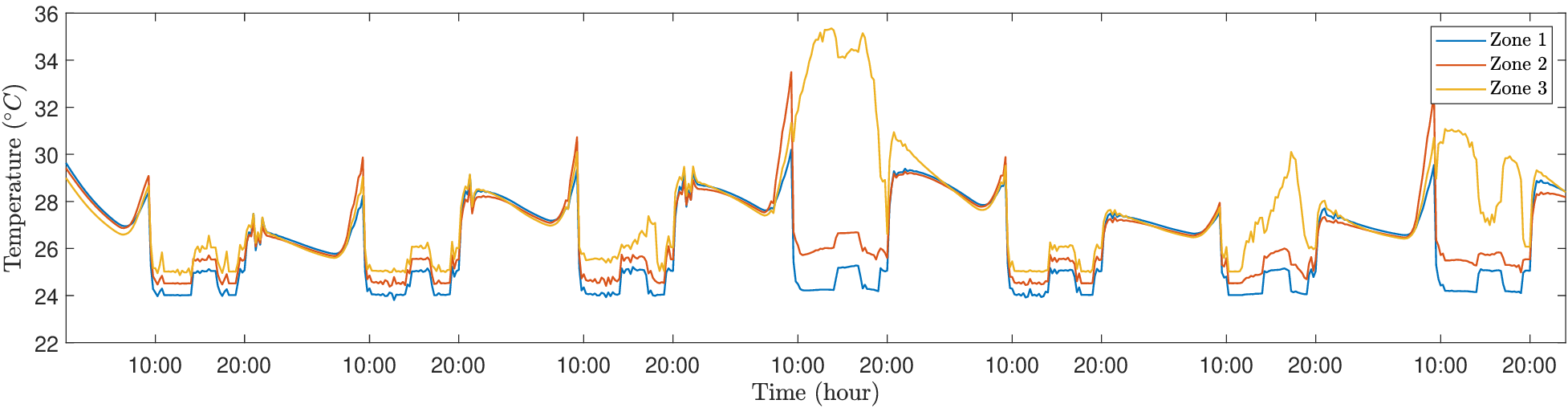}   
		\end{minipage}
		\label{centralized}
	}
	\subfigure[Decentralized strategy.] 
	{
		\begin{minipage}[t]{0.9\linewidth}
			\centering      
			\includegraphics[width=1\textwidth]{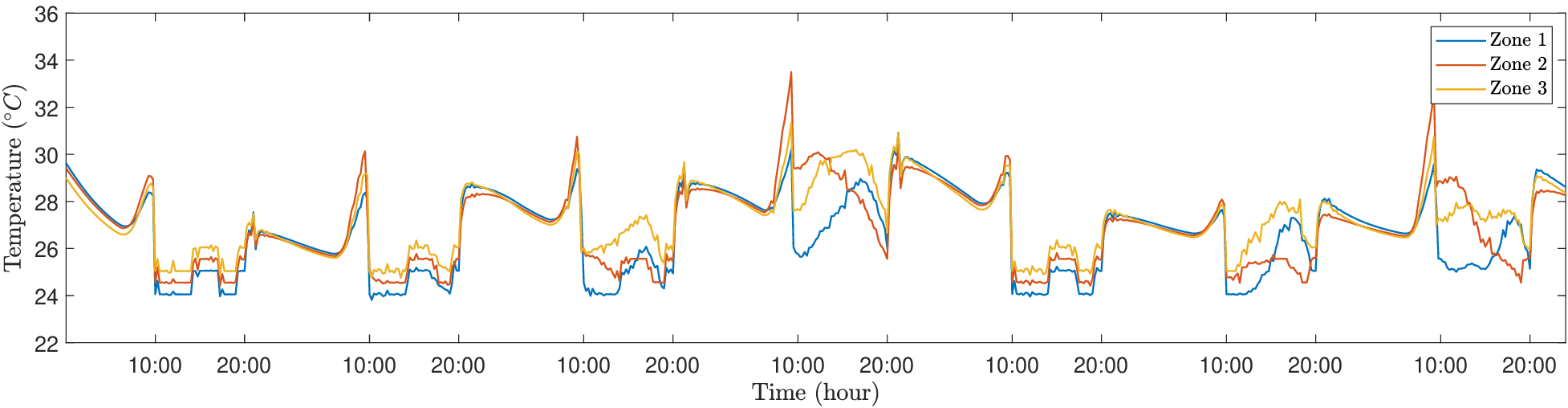}   
		\end{minipage}
		\label{decentralized}
	}
	\subfigure[Distributed strategy.] 
	{
		\begin{minipage}[t]{0.9\linewidth}
			\centering      
			\includegraphics[width=1\textwidth]{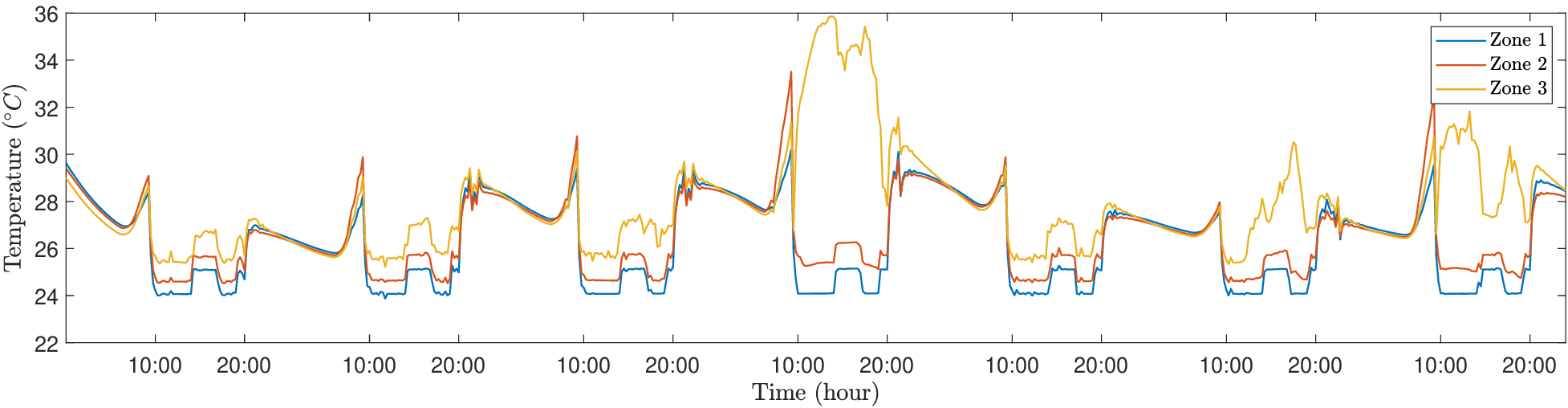}   
		\end{minipage}
		\label{distributed}
	}
	\caption{Indoor temperature for small-scale scenarios.} 
	\label{temperatures_centralialed}  
\end{figure*}

Set $\theta$ in the centralized scheme to $\theta_1=1,\theta_2=0.1,\theta_3=0.01$.
The temperature change during the 7-day period under the centralized strategy is shown in Fig.\ref{centralized}. When there is sufficient energy supply, the temperatures in all three zones meet the control requirements well. When there is insufficient energy supply, Zone\:1, which has the highest priority, is met first.

The temperature change of the decentralized strategy is shown in Fig. \ref{decentralized}. When the energy supply is sufficient, the decentralized strategy provides a similar control effect to that provided by the centralized strategy. When the energy supply is insufficient, the decentralized strategy is not able to distribute energy well, and thus some rooms have excess energy while others lack energy, which eventually leads to a worse control effect compared with the centralized strategy.

The temperature change curves obtained under the distributed strategy are shown in Fig. \ref{distributed}. Regardless of whether the energy supply is sufficient or not, the distributed strategy proposed in this paper ensures the comfort of the zone with the highest priority first, i.e., it can be observed that the temperature of Zone\:1 (highest priority) in Fig. \ref{distributed} is always maintained near the comfortable temperature range, and then the energy supply of other rooms is considered in the order of priority.
 
\begin{figure*}[htbp]
	\centering    
	\subfigure[First priority.] 
	{
		\begin{minipage}[t]{0.28\linewidth}
			\centering          
			\includegraphics[width=1\textwidth]{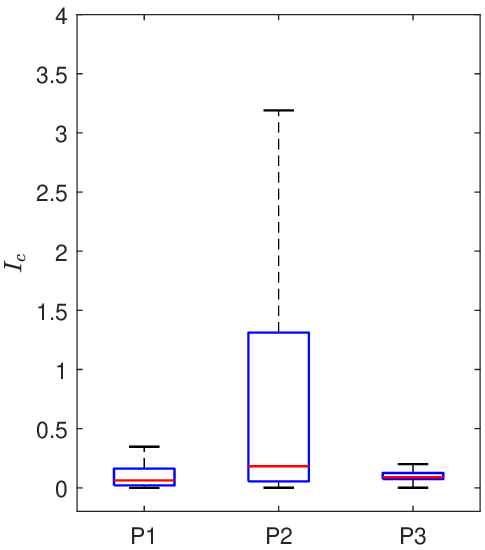}   
		\end{minipage}
		\label{3zones_centralized}
	}
	\subfigure[Second priority.] 
	{
		\begin{minipage}[t]{0.28\linewidth}
			\centering      
			\includegraphics[width=1\textwidth]{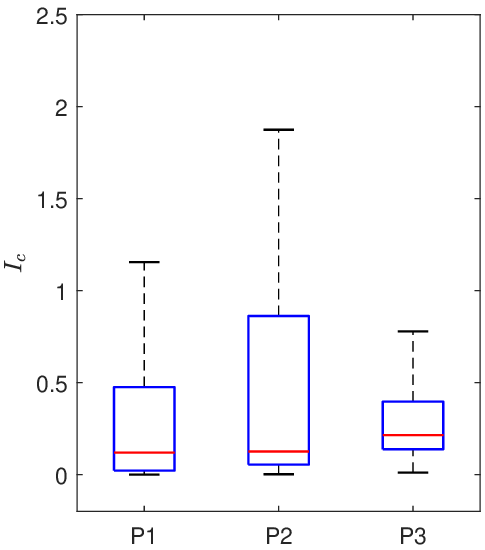}   
		\end{minipage}
		\label{3zones_decentralized}
	}
	\subfigure[Third priority.] 
	{
		\begin{minipage}[t]{0.28\linewidth}
			\centering      
			\includegraphics[width=1\textwidth]{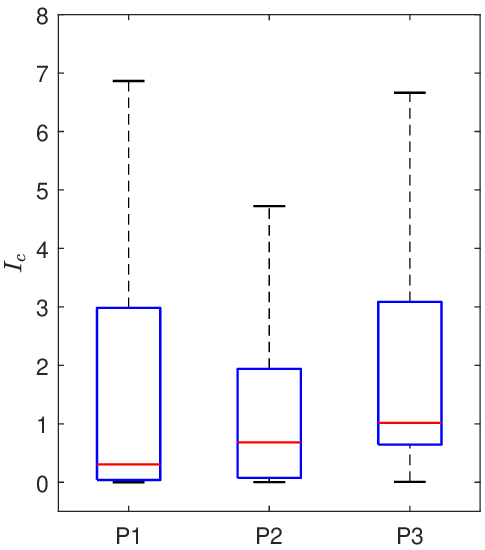}   
		\end{minipage}
		\label{3zones_distributed}
	}
	\caption{Comfort index for the three schemes in the small-scale scenario (P1: Centralized, P2: Decentralized, P3: Distributed).} 
	\label{index3zones}  
\end{figure*}

Comfort data is collected every fifteen minutes during the 7-day simulation. Fig. \ref{index3zones} shows the distribution and variability of the comfort index for the three strategies in the 7-day simulation. It can be seen that in both the centralized MPC and distributed MPC schemes, Zone\:1, with the highest priority, is able to maintain a suitable temperature range, and Zone\:2, with the second highest priority, is able to satisfy a certain degree of energy supply with sufficient supply from Zone\:1. This shows that the performance of the proposed distributed strategy is very close to that of the centralized strategy. 

Table \ref{un1} and Table \ref{cool1} show the comfort index and the energy rate for different priority zones in the small-scale scenario. Our distributed scheme has the lowest $I_{c1}$ and the largest energy rate in the highest priority zone (compared with the centralized and decentralized scheme), which shows that our solutions are most comfortable for the highest priority zone.

\begin{table}[htbp]
	\centering
	\caption{$I_c$ for different priority zones in the small-scale scenario 
	}
	\label{un1}
	\small 
	\begin{tabular}{c|c|c|c|c}
		\hline 
		&$I_{c1}$&$I_{c2}$ &$I_{c3}$ &$I_{co}$\\
		\hline 
		Centralized& 0.1081 & 0.3620 & 1.9402&0.2499\\
		\hline  
		Decentralized& 0.8230 & 0.9299 & 1.1555&0.8815\\
		\hline  
		Distributed& 0.1059 & 0.4047 &2.0223 &0.2617\\
		\hline  
	\end{tabular}
\end{table}

\begin{table}[htbp]
	\centering
	\caption{Energy rate (unit: $W$) for different priority zones in the small-scale scenario 
	}
	\label{cool1}
	\small 
	\begin{tabular}{c|c|c|c}
		\hline 
		&first priority&second priority&third priority \\
		\hline 
		Centralized& 746.1870 & 689.5958 & 615.9371\\
		\hline  
		Decentralized& 677.7220 & 643.6687 & 695.7191\\
		\hline  
		Distributed& 781.9922 & 725.3697&595.4177 \\
		\hline  
	\end{tabular}
\end{table}

The computational costs in the three schemes are shown in Table \ref{time3}. The distributed computational cost is close to the decentralized one, and both are better than the centralized one. In this small-scale scenario, the longest time of sequential computation on one PC is taken to calculate the computational cost of decentralized and distributed schemes. In fact, we can implement parallel computation using parallel pools in MATLAB instead of sequential computation. However, it takes a lot of time to start the parallel pool and allocate worker threads, and it is not worthwhile to perform parallel computation with only three subsystems. Therefore, the longest time for sequential computation is used instead of the time for parallel computation in the small-scale case. In the following large-scale case (36 zones), we use the parallel computation time on one PC in the decentralized and distributed schemes.

\begin{table}[htbp]
	\centering
	\caption{Computational Cost Comparison}
	\label{time3}
	\small 
	\begin{tabular}{ccc}
		\hline 
		Centralized&Decentralized&Distributed\\
		\hline  
		125.5869s&77.48s&78.77s\\
		\hline
	\end{tabular}\\
        \footnotesize{$*$ The longest time of sequential computation is employed for decentralized and distributed computational costs on one PC.}\\
\end{table}

\subsection{Test results in the large-scale scenario}
\begin{figure*}[htbp]
	\centering    
	\subfigure[Centralized strategy.] 
	{
		\begin{minipage}[t]{0.9\linewidth}
			\centering          
			\includegraphics[width=1\textwidth]{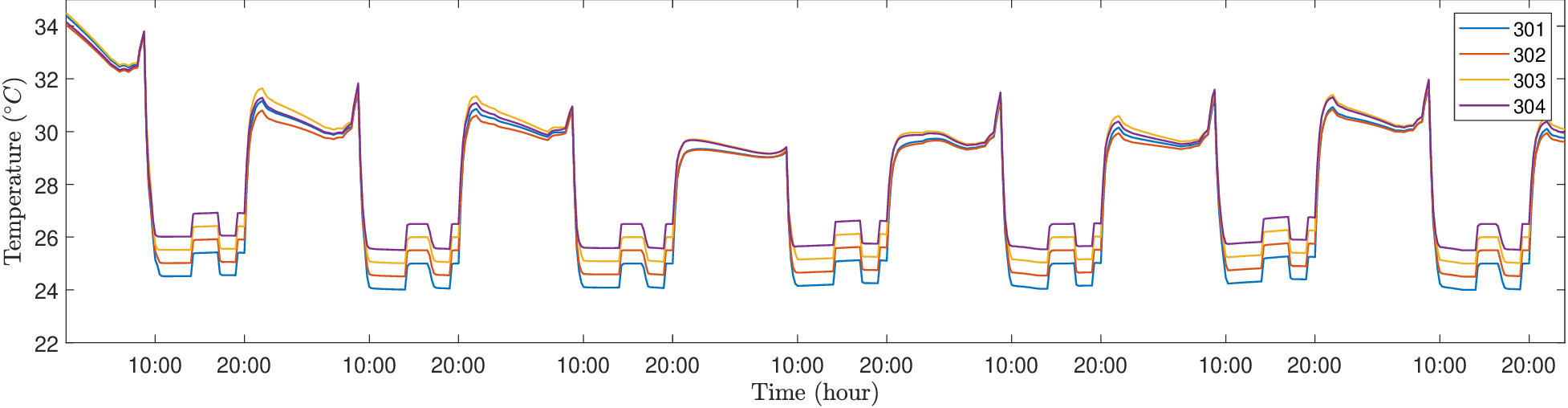}   
		\end{minipage}
		\label{36zones_centralized1}
	}
	\subfigure[Decentralized strategy.] 
	{
		\begin{minipage}[t]{0.9\linewidth}
			\centering      
			\includegraphics[width=1\textwidth]{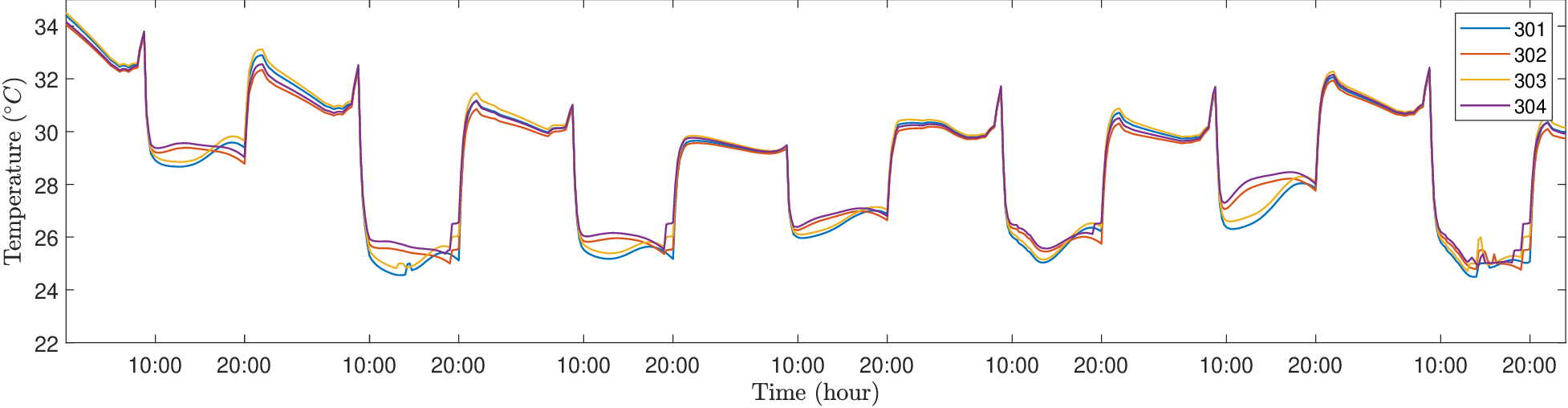}   
		\end{minipage}
		\label{36zones_decentralized1}
	}
	\subfigure[Distributed strategy.] 
	{
		\begin{minipage}[t]{0.9\linewidth}
			\centering      
			\includegraphics[width=1\textwidth]{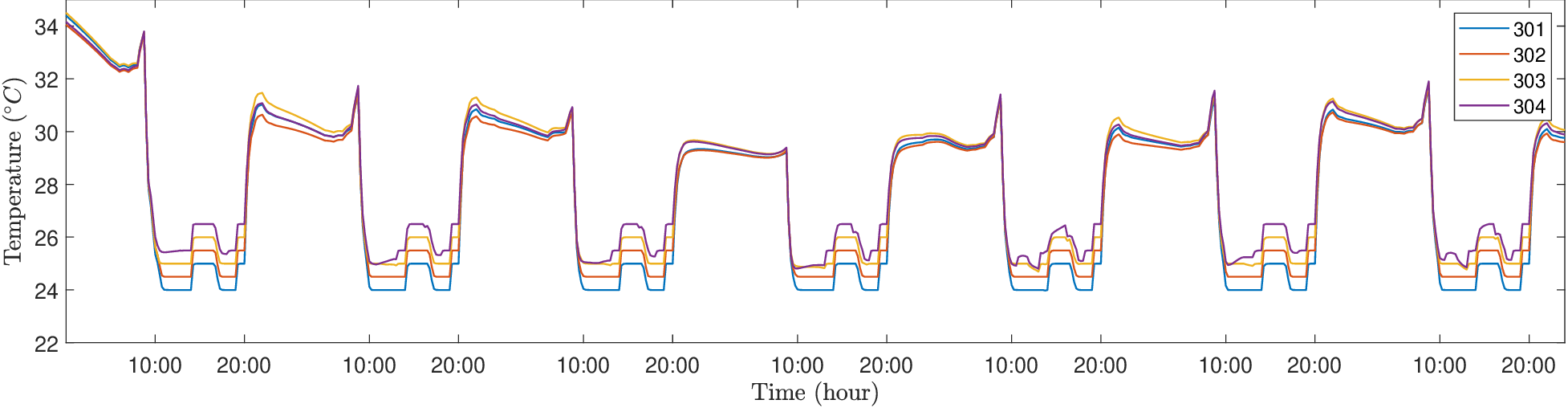}   
		\end{minipage}
		\label{36zones_distributed1}
	}
	\caption{Indoor temperature for zones with second priority (floor 3).} 
	\label{temperatures_centralialed36}  
\end{figure*}

Set $\theta$ in the centralized scheme to be $\theta_1=1,\theta_2=0.1,\theta_3=0.01$. The temperature curves of zones with second priority under centralized, decentralized and distributed MPC strategies are shown in Fig. \ref{temperatures_centralialed36}. Similar situations to that of the small-scale scenario can be observed. The centralized MPC strategy is able to supply more sufficient energy to the zones with higher-priority by adjusting the weights. The decentralized MPC strategy has inflexible energy scheduling and is not able to satisfy the energy supply in the high-priority zones. The distributed MPC strategy has similar control performance as the centralized MPC strategy, and it is able to satisfy the energy supply of the high priority zones even if the energy supply system does not provide enough energy for the whole system.

\begin{figure*}[htbp]
	\centering    
	\subfigure[First priority.] 
	{
		\begin{minipage}[t]{0.28\linewidth}
			\centering          
			\includegraphics[width=1\textwidth]{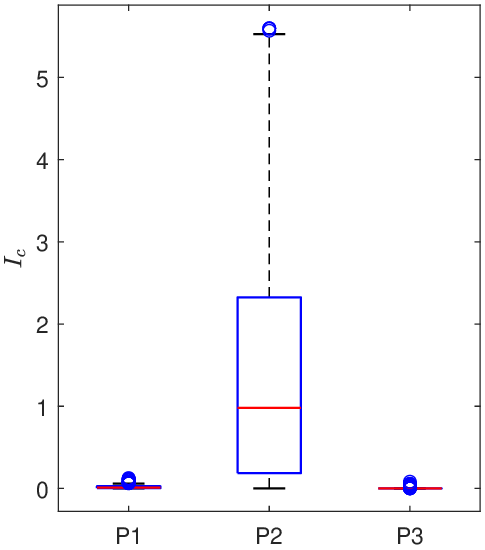}   
		\end{minipage}
		\label{36zones_centralized}
	}
	\subfigure[Second priority.] 
	{
		\begin{minipage}[t]{0.28\linewidth}
			\centering      
			\includegraphics[width=1\textwidth]{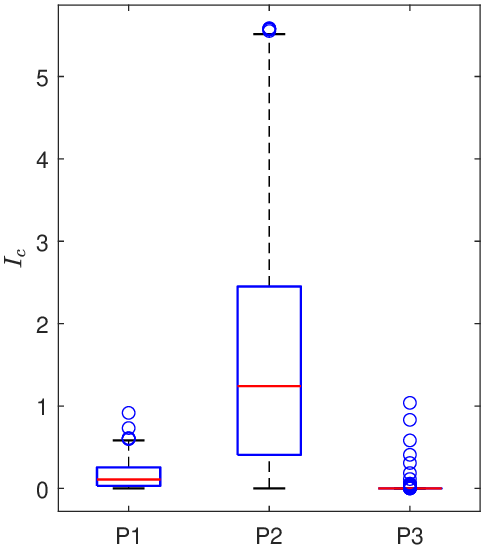}   
		\end{minipage}
		\label{36zones_decentralized}
	}
	\subfigure[Third priority.] 
	{
		\begin{minipage}[t]{0.28\linewidth}
			\centering      
			\includegraphics[width=1\textwidth]{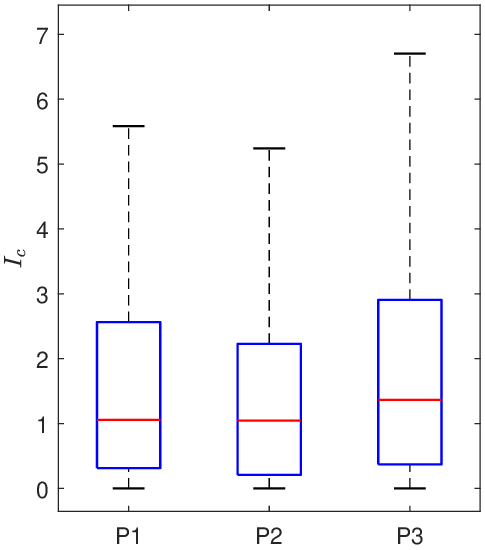}   
		\end{minipage}
		\label{36zones_distributed}
	}
	\caption{Comfort index for the three schemes in the large-scale scenario (P1: Centralized, P2: Decentralized, P3: Distributed).} 
	\label{index36zones}  
\end{figure*}

Comfort data is collected every fifteen minutes during the 7-day simulation. The distribution and variability of the comfort index for the three strategies simulated for 7 days is shown in Fig. \ref{index36zones}. The decentralized scheme can only make the temperature comfortable in the zones with high priority when the energy supply is sufficient and cannot cope with the scenario of insufficient energy supply. Both the centralized scheme and the distributed scheme are able to make the zones with high supply priority as comfortable as possible. 

Table \ref{un} and \ref{cool2} show the comfort index and the energy rate for different priority zones in the large-scale scenario, respectively. Our distributed scheme has the lowest $I_{c1}$ and the largest energy rate in the highest priority zone (compared to the centralized and decentralized schemes), which indicates that our distribution scheme is the most reasonable. 

\begin{table}[htbp]
	\centering
	\caption{$I_c$ for different priority zones in the large-scale scenario }
	\label{un}
	\small 
	\begin{tabular}{c|c|c|c|c}
		\hline 
		&$I_{c1}$&$I_{c2}$ &$I_{c3}$ &$I_{co}$\\
		\hline 
		Centralized& 0.0173 & 0.1613 & 1.5807&0.1670\\
		\hline  
		Decentralized& 1.2633 & 1.4279 & 1.2662&1.3476\\
		\hline  
		Distributed& 0.0010 & 0.0041 &1.7107 &0.1711\\
		\hline  
	\end{tabular}
\end{table}

\begin{table}[htbp]
	\centering
	\caption{Energy rate (unit: $W$) in the large-scale scenario 
	}
	\label{cool2}
	\small 
	\begin{tabular}{c|c|c|c}
		\hline 
		&first priority&second priority&third priority \\
		\hline 
		Centralized& 778.1782 & 855.7001 & 711.3001\\
		\hline  
		Decentralized& 671.0186 & 738.8950 & 742.2782\\
		\hline  
		Distributed& 788.9912 & 880.5024&705.3784 \\
		\hline  
	\end{tabular}
\end{table}

The computational cost of the three schemes is shown in Table \ref{time}. The size of the optimization problem to be solved by the centralized controller is related to the number of subsystems, and when the number of subsystems is relatively large, the solution is less efficient, so the computational cost of the centralized strategy is the highest among the three strategies. The size of the optimization problem to be solved by the decentralized strategy is close to that of the distributed strategy, but the computational cost of the distributed strategy is slightly higher than that of the decentralized strategy due to the exchange of information between subsystems. Since the distributed strategy performs well and the optimization problem to be solved by each sub-controller does not change as the number of subsystems increases, the distributed strategy is more suitable for scaling up to large-scale systems. It is noted that the decentralized and distributed computational costs are obtained using parallel pools in MATLAB on one PC. Thus, the decentralized and distributed computational costs also include the time to start parallel pools and allocate worker threads. In addition, due to the limited memory and CPU cores on one PC, it takes longer to implement parallel computation on one PC than on multiple PCs. However, the distributed computational cost is still much better than the centralized solution, even on one PC.
\begin{table}[htbp]
	\centering
	\caption{Computational Cost Comparison}
	\label{time}
	\small 
	\begin{tabular}{ccc}
		\hline 
		Centralized&Decentralized&Distributed\\
		\hline  
		719.22s&313.43s&326.90s\\
		\hline
	\end{tabular}\\
        \footnotesize{$*$ The decentralized and distributed computational costs are obtained using parallel pools in MATLAB on one PC.}\\
\end{table}

\section{Conclusions}
In this paper, a priority-based energy distribution scheme is developed that rationally distributes energy based on priority order. For different operation scenarios, a one-to-one priority strategy and a multi-to-one priority strategy based on distributed MPC are proposed. The one-to-one priority strategy refers to the case that a single subsystem corresponds to a particular priority, and the multi-to-one strategy refers to the situation that multiple subusystems correspond to the same priority level. By exploiting the property of MPC, i.e., obtaining all the solutions in the control horizon, the subsystems can perform optimization operations in parallel.

Simulation experiments of the proposed strategy have been carried out in a three-zone building and a 36-zone building, respectively.  The experimental results show that the developed solution could satisfy the urgent energy supply in some specific zones, and the performance is close to that of the centralized scheme.

\section*{ACKNOWLEDGMENT}
This work was supported in part by the National Natural Science Foundation of China under Grant 62173113, and in part by the Science and Technology Innovation Committee of Shenzhen Municipality under Grant GXWD20231129101652001, and in part by Natural Science Foundation of Guangdong Province of China under Grant 2022A1515011584.

\bibliographystyle{ieeetr}
\bibliography{reference} 

\begin{IEEEbiography}[{\includegraphics[width=1in,height=1.25in,clip,keepaspectratio]{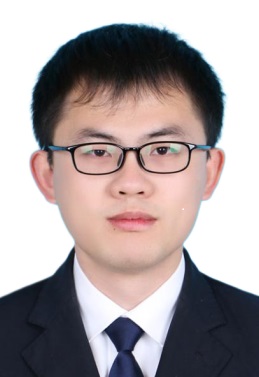}}]{Hongyi Li}
	received her B.S. degree in Automation and M.S. degree in Control Science and Engineering from Harbin Institute of Technology, China, in 2020 and 2023, respectively. He is currently working toward the Ph.D. degree in Control Science and Engineering from Harbin Institute of Technology, Shenzhen, China. His research interests include piecewise linear approximation and building energy conservation. \end{IEEEbiography}

\begin{IEEEbiography}[{\includegraphics[width=1in,height=1.25in,clip,keepaspectratio]{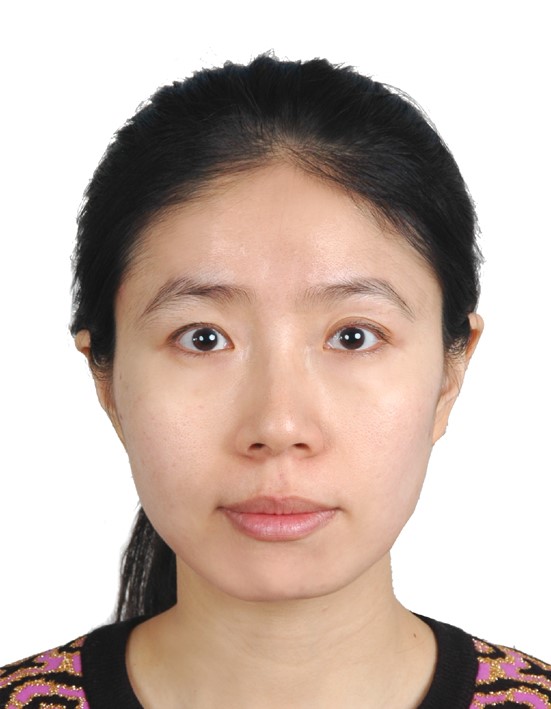}}]{Jun Xu}
	 received her B.S. degree in Control Science and Engineering from Harbin Institute of Technology, Harbin, China, in 2005 and Ph.D. degree in Control Science and Engineering from Tsinghua University, China, in 2010. Currently, she is an associate professor in School of Mechanical Engineering and Automation, Harbin Institute of Technology, Shenzhen, China. Her research interests include piecewise linear functions and their applications in machine learning, nonlinear system identification and control.\end{IEEEbiography}

\begin{IEEEbiography}[{\includegraphics[width=1in,height=1.25in,clip,keepaspectratio]{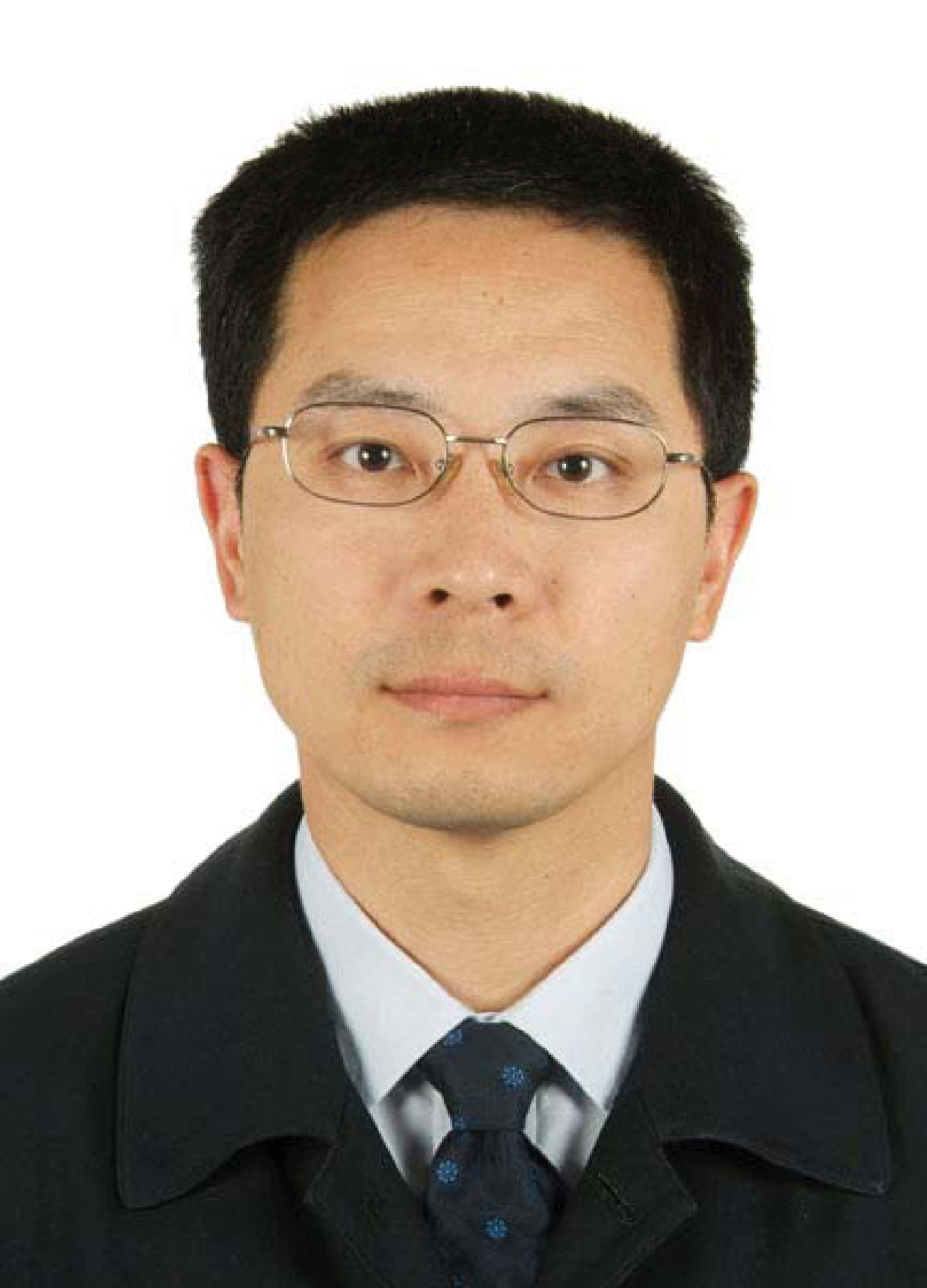}}]{Qianchuan Zhao}
	 (Senior Member, IEEE) received the B.E. degree in automatic control and the B.S. degree in applied mathematics and the M.S. and Ph.D. degrees in control theory and its applications from Tsinghua University, Beijing, China, in 1992 and 1996, respectively. He was a Visiting Scholar with Carnegie Mellon University, Pittsburgh, PA, USA, in 2000; and Havard University, Cambridge, MA, USA, in 2002. He was a Visiting Professor with Cornell University, Ithaca, NY, USA, in 2006. He is currently Professor and the Director of Center for Intelligent and Networked Systems, Department of Automation, Tsinghua University. He has published more than 100 research papers in peer-reviewed journals and conferences. His research interests include the control and optimization of complex networked systems with applications in smart buildings, and manufacturing automation. Prof. Zhao received the 2009 and 2018 China National Nature Science Awards and the 2014 National Science Foundation for Distinguished Young Scholars of China. He serves as the Chair of the Technical Committee on smart buildings of the IEEE Robotics and Automation Society. He is currently the Editor-in-Chief of the Results in Control and Optimization (RICO), an Editor of the IEEE TRANSACTIONS ON AUTOMATION SCIENCE AND ENGINEERING, and an Associate Editor of the Journal of Optimization Theory and Applications.\end{IEEEbiography}

\end{document}